\documentclass[aps,pre,showpacs,floatfix]{revtex4}
\usepackage{graphicx,amsfonts,psfrag}

\begin{document}

\title{Fragmentation of a circular disc by Impact on a Frictionless plate}
\author{Bhupalendra Behera$^{1,3}$,   
 Ferenc Kun${}^2$, Sean McNamara${}^1$, and Hans J.\ Herrmann${}^1$}
\affiliation{${}^1$Institut f\"ur Computeranwendungen (ICA1), Universit\"at
  Stuttgart, 70569 Stuttgart, Germany \\
  ${}^2$Department of Theoretical Physics, University of
  Debrecen, H-4010 Debrecen, P.O.Box: 5, Hungary\\
  ${}^3$Indian Institute of Technology, Department of Materials 
  and Metallurgical Engineering, Kanpur, India
}
\email{bhupalendra@yahoo.com}

\date{\today}

\pacs{64.60.-i, 64.60.Ak, 46.30.Nz}

\begin{abstract}

The break-up of a two-dimensional circular disc by normal and oblique
impact on a hard frictionless plate is investigated by molecular
dynamics simulations. The disc is composed of numerous unbreakable
randomly shaped convex polygons connected together by simple elastic
beams that break when bent or stretched beyond a certain limit.
It is found that for both normal and oblique impacts
the crack patterns are the same and depend solely on the normal
component of  the impact velocity. Analysing the pattern of breakage,
amount
of damage, fragment masses and velocities, we show the existence of a
critical velocity which separates two regimes of the impact process:
below the critical point only a damage cone is formed at the impact
site {\it (damage)}, cleaving of the particle occurs at the critical
point, while above the critical velocity the disc breaks into several
pieces {\it (fragmentation)}. In the limit of very high impact
velocities the disc suffers complete disintegration {\it (shattering)}
into many small fragments. In agreement with experimental results,
fragment masses are found to follow the Gates-Gaudin-Schuhmann
distribution (power law) with an exponent independent of
the velocity and angle of impact.
The velocity distribution of fragments exhibit an
interesting anomalous scaling behavior when changing the impact
velocity and the size of the disc.
\end{abstract}
\maketitle

\section{Introduction}
\label{intro}
The strength and break-up of agglomerates composed of smaller sized
primary particles is of particular importance for the storage
and handling of materials in process industries such as
pharmaceuticals, chemicals, fertilizers, detergent, and food
industries. In industrial processes agglomerates often collide
with each other and with the hard walls of the equipment resulting in a
size reduction, which is desired or not depending on the
type of the process. The strength of agglomerates has to be
characterized for the design of operating conditions in industrial
processes such as milling, tabletting, mixing, and transport in
pneumatic conveying. Another
important class of agglomerates are the so-called particle compounds,
which are the combination of various sized particles embedded in a
cementous matrix. The different types of engineering agglomerates and
building materials like concretes are some examples of particle
compounds. It is of high industrial importance to recycle these
particle compounds in order to use the valuable aggregates.
The design and optimization of the liberation process of
aggregates from the matrix material requires a detailed knowledge of
the strength and break-up of compounds.

For the understanding of the strength and
break-up process,  the study of simple systems like
spherical particles is essential. During the last decades several
experimental and theoretical studies 
have been performed to understand the break-up of spherical bodies
arising due to impact. The crack pattern of sand-cement spheres by a
free fall impact was studied in Ref.\ 
\cite{fra_pattern}, which reports observations of   meridian cracks,  that divide
the sphere into two nearly equal parts, and 
oblique cracks, which are straight like median cracks, but cut the
sphere into two unequal pieces. The  fracture of glass and plaster
spheres by free fall impact and double impact (dynamic loading between
two hard plates) have been carried out recently
\cite{powder,chau}. It was found that at the lowest impact velocities
hertzian cone cracks (formed from a surface ring crack) 
are developed, whereas, at high velocities,  oblique cracks
propagate  before meridian cracks form \cite{fra_glass}.
This finding differs from the experimental results of
Ref.\ \cite{fra_pattern}, where it was found that with increasing
impact energy, the number of meridian planes increases and oblique
cracks start to develop. 

Due to the high speed and violent nature of the break-up process,
observations are usually restricted to the final state of impact
experiments, where information 
has to be extracted from the remaining pieces of the body. Hence,
computer simulation of models of agglomerate break-up is an
indispensable tool in this field. Simulations of realistic models
provide a deeper inside into the break-up process and can even
complement the experimental findings directly supporting the design
of industrial processing of these materials. Analytic approaches have
limited capabilities in this field since they cannot capture the
disordered microstructure of the material. 

The finite element approach and the discrete element modeling have been
successfully applied to describe the stress field, crack propagation, 
and fragment formation in impacting spherical particles
\cite{agglomerate,bk,imp_agglo,poto,simu,tsoungui,imp_ang,poschel1,poschel2}.
Recent simulations of ball impact revealed two types
of crack patterns: oblique cracks radiating from impact point, and
secondary cracks perpendicular to the oblique ones. 
In the framework of the
discrete element method it was clarified that depending on the impact
velocity the result of the break-up process can be localized damage
around the contact zone, fragmentation, or shattering. The evolution of several characteristic quantities of
the break-up process when increasing the impact velocity were monitored
and analyzed in normal and oblique impact
\cite{agglomerate,bk,imp_agglo,poto,imp_ang}. 

From a more general point of view, the break-up of agglomerates presents
an important class of fragmentation phenomena which is ubiquitous in
everyday life and concerns a wide range of phenomena in science and
technology. 
In general, when an object is subjected to shock or stress it 
will break up into smaller pieces.  The length scales 
involved in this process range from the collisional evolution of asteroids 
\cite{asteroids} to the degradation of materials comprised of small 
agglomerates \cite{agglomerate,bk} employed in the process industries
as summarized above. 
There are also many geological examples associated with the use of
explosives for mining and oil shale industry, coal heaps, etc. 
A wide variety of experiments
\cite{self,composite,instable,ice,lab,glassplate}  
and simulations \cite{agglomerate,bk,discrete,transition,twodisc,granulate,diehl,univer,branching,aspect,droplet,britt,poto,simu} 
revealed that the fragment mass distribution is a power 
law  except for very large fragment sizes. The
exponents in the power law region were found experimentally to be
between 1.35 and 2.6 depending on the effective dimensionality of the
system \cite{asteroids,fra_pattern,lab,breakup}. 
Recent studies revealed that power law distributions
arise in fragmentation phenomena due to an underlying phase
transition \cite{instable,transition,univer}. 
However, most of the data reported in the 
literature is concerned with the general behavior of fragmentation 
processes. There is much less literature  where the 
propagation and orientation of cracks are discussed. 

In the present paper we study the normal and oblique impact of a
circular brittle particle  on a hard frictionless plate, varying the
impact velocity and impact angle in a broad range. 
The particle is composed of numerous unbreakable, undeformable,
randomly shaped polygons which are bonded together by elastic
beams. The bonds between the polygons can be broken according to a
physical breaking rule, which takes into account the stretching and
bending of the connections.
Based on simulations of the model, we performed a detailed study of the
failure evolution at different impact velocities
and of the nature of the crack propagation during the fragmentation
process, and compared the results with experiments 
\cite{fra_pattern,powder,lab,ice,fra_glass}. In the analysis of the
simulation data, we profit from recent theoretical results of general
studies of fragmentation processes. 
We observed that for both normal and oblique impacts, 
the crack patterns are the same and depend solely on the normal
component of  the impact velocity. 
Studying the crack patterns, amount of damage, fragment masses, and
velocities, 
we provide a quantitative foundation of the concept of 
damage, fragmentation, and shattering in ball impact, which was
introduced recently on a more qualitative basis \cite{agglomerate}. We
show the 
existence of a critical impact velocity $v_c$ which distinguishes two
regimes of the impact process, {\it i.e.} below the critical velocity
damage mainly occurs in a conical region around the impact site with a
large residue, however, above $v_c$ an ensemble of oblique cracks
develop and the disc breaks up into pieces. In agreement with
experimental results, fragment masses are found
to follow the Gates-Gaudin-Schuhmann distribution (power law)
\cite{kelly} with an exponent independent of the velocity and
angle of impact. The velocity distribution of fragments exhibit an
interesting anomalous scaling behavior when changing the impact
velocity and the size of the disc.

An important application of our results, besides the ones mentioned at
the beginning, is to the optimization and control 
of tumbling mill performance. These questions are of utmost 
practical importance as they have a tremendous influence on power draft, 
wear of the balls and liners and breakage characteristics of the grinding
materials. During the cataracting motion where the charge material inside
a  mill follows a parabolic path \cite{data}, most of the materials 
are ground as  hard balls fall back onto them. There is particular interest
in the net energy required to achieve a certain size reduction and the
the energy distribution of the fragments during the grinding process.
 The efficiency of the mills
could be controlled if the breakage characteristics of the grinding 
materials were better understood. Our current work can provide some valuable
information for  the modernization of the mill design.  

\section{Model}
\label{model}
In order to study fragmentation of granular solids, 
we performed molecular dynamic (MD)
simulations in two dimensions. To better capture the complex
structure of a real solid, we used randomly generated convex polygons
that interact with each other elastically. The model consists of three
major parts, namely, the construction of a Voronoi cellular structure, the
introduction of the elastic behavior, and finally the breaking of the solid.
This section gives a detailed overview of these three steps. 

In order to take into account the complex structure of the granular solid, 
we use randomly generated convex polygons, {\it i.e.} we divide the solid into
grains by a Voronoi cellular structure. The Voronoi construction 
is a random tessellation of the plane into convex polygons. 
This is obtained by putting  a random set of points onto the plane 
and then assigning to each point that part of the plane which is nearer 
to it than to any other point. One advantage of the
Voronoi tessellation is that the number of neighbors of each polygon is
limited which makes the computer code faster and allows us to simulate 
larger systems. In our case,  the initial configuration of the polygons was constructed 
using a vectorizable random lattice, which is Voronoi construction
 with slightly reduced disorder \cite{rand_lattic}.  First, the Voronoi tessellation of a
square is performed, and then a circular disc  with smooth
surface is cut out.

In the model the polygons are rigid bodies.
They are neither breakable nor deformable, but they can overlap 
when pressed against each other. This overlap represents local deformations 
of the grains. Usually the overlapping polygons have two intersection 
points which define the contact line. 
In order to simulate the elastic contact force, we introduce
a repulsive force between touching polygons. This force is proportional to
the overlapping area $A$ divided by a characteristic length $L_c$
$(\frac{1}{L_c}= \frac{1}{2}[\frac{1}{r_i}+\frac{1}{r_j}]$, where
$r_i$, $r_j$ are the radii of circles of the same area as the polygons).
 The direction of the elastic or normal 
force is perpendicular to the contact line of the polygons.
The complete form of the normal force  contains an elastic 
and damping contribution, whereas the tangential component is responsible 
for the friction.

Again, to bond the particles together it is necessary to introduce a 
cohesive force between neighboring polygons. For this purpose we 
introduce beams. The centers of mass of neighboring
polygons are joined together with elastic beams that exert an attractive,
restoring force but can break in order to model the
fragmentation of the solid.  Because of the randomness contained in the 
Voronoi tessellation, the lattice of beams is also random. The length, the
cross-section and the moment of inertia of each beam are determined
by the initial configuration of the polygons. The Young's
modulus of the beams and of the particles are considered to be
independent of each other. The beams break according to a
physical breaking rule, which takes into account the stretching and
bending of the connection. 
The surface of the grains where beams are broken represent cracks.
 The energy stored in the broken beams represents the energy needed to create
these new crack surfaces inside the solid. 

In order to simulate the break-up of the disc due to impact with a
hard plate, a repulsive force is introduced between the plate and
those polygons of the disc which have overlap with the plate. This
repulsive force is proportional to the overlap area, similarly to the
polygon-polygon contacts but with a higher stiffness value. The
contact force of the disc and the plate has vertical direction,
tangential component like friction is excluded in the present study. 

The time evolution of the system is obtained by numerically solving
Newton's equations of motion of the individual polygons (Molecular
Dynamics).  For the solution of the equations we use a Gear
Predictor-Corrector scheme of fifth order, which means that we have to
keep track of the coordinates and all their derivatives up to fifth
order. The breaking criterion of beams is evaluated in each iteration
time step and those beams which fulfill the condition are removed
from the calculations. The simulation stops after no beams break
during a certain number of
time steps.   
Previously this model has been applied to study fragmentation of
solids in various experimental situations
\cite{discrete,transition,twodisc,granulate,proj}. For more details of the
simulation technique see Ref.\ \cite{discrete}. 

\section{Crack pattern}
\label{crack}
\begin{figure}
\begin{tabular}{ccc}
\includegraphics[width=0.3\textwidth]{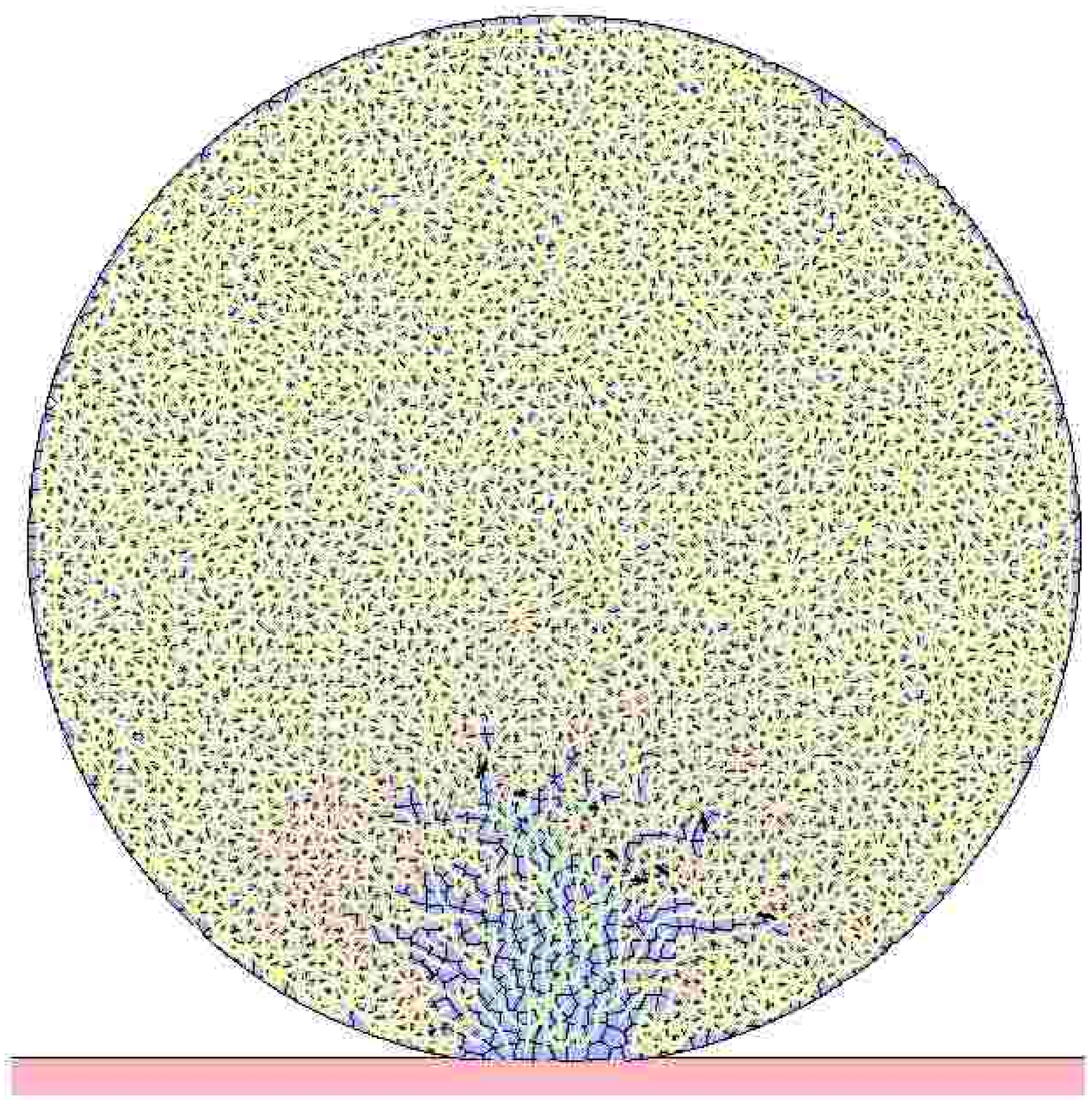}&
\includegraphics[width=0.3\textwidth]{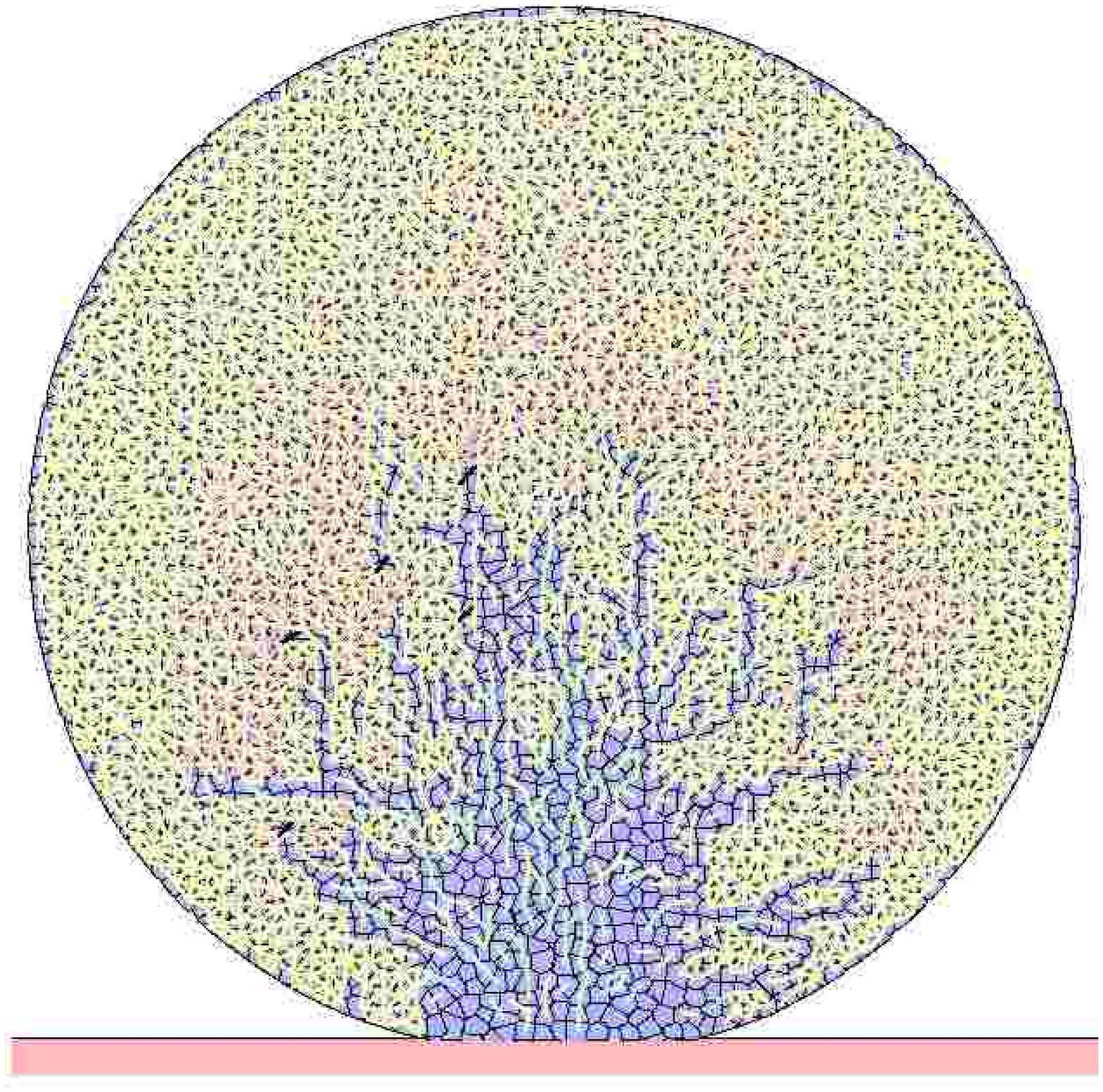}&
\includegraphics[width=0.3\textwidth]{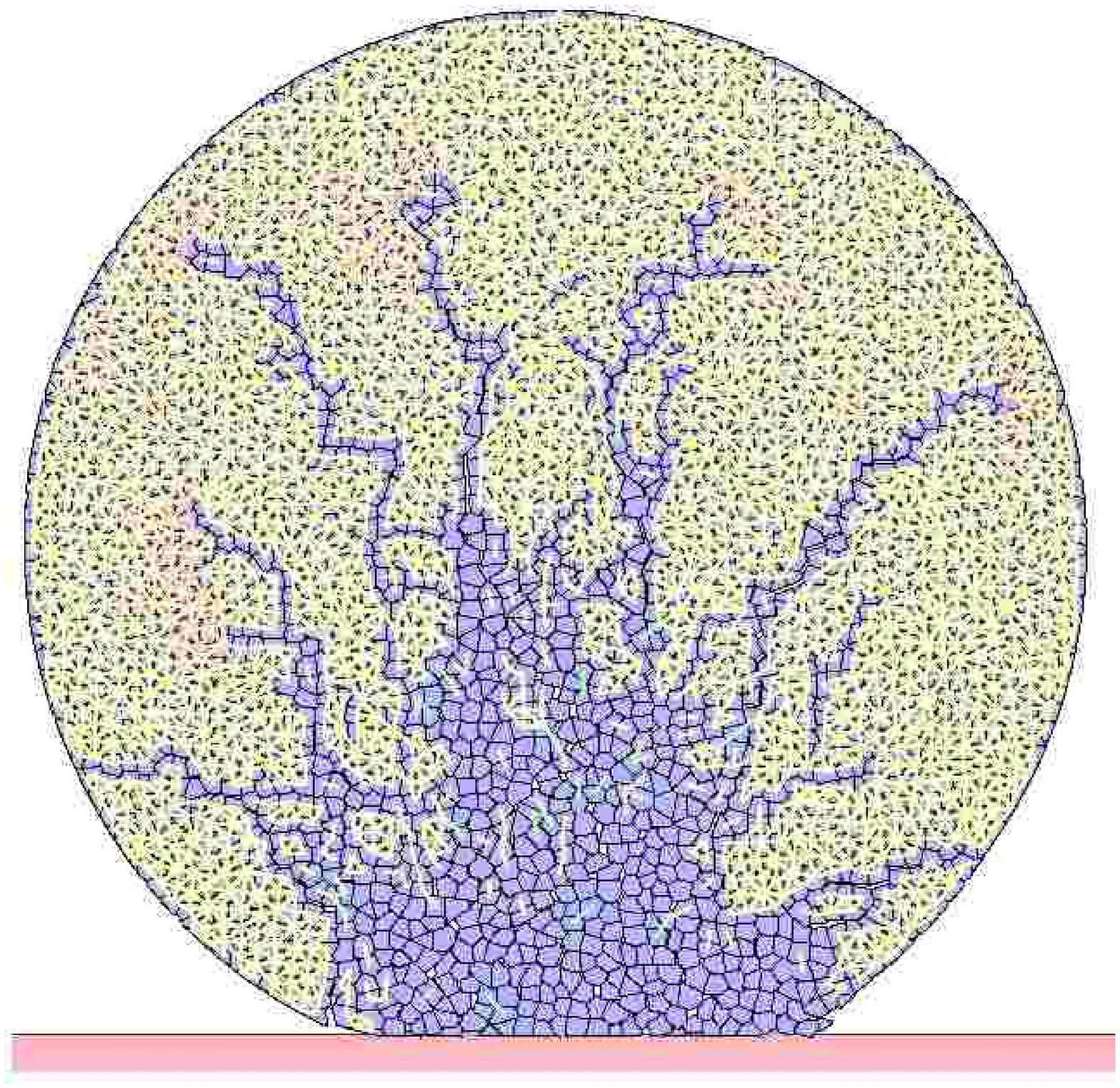}\\
(a)&(b)&(c)\\
\end{tabular}
\begin{tabular}{cc}
\includegraphics[width=0.4\textwidth]{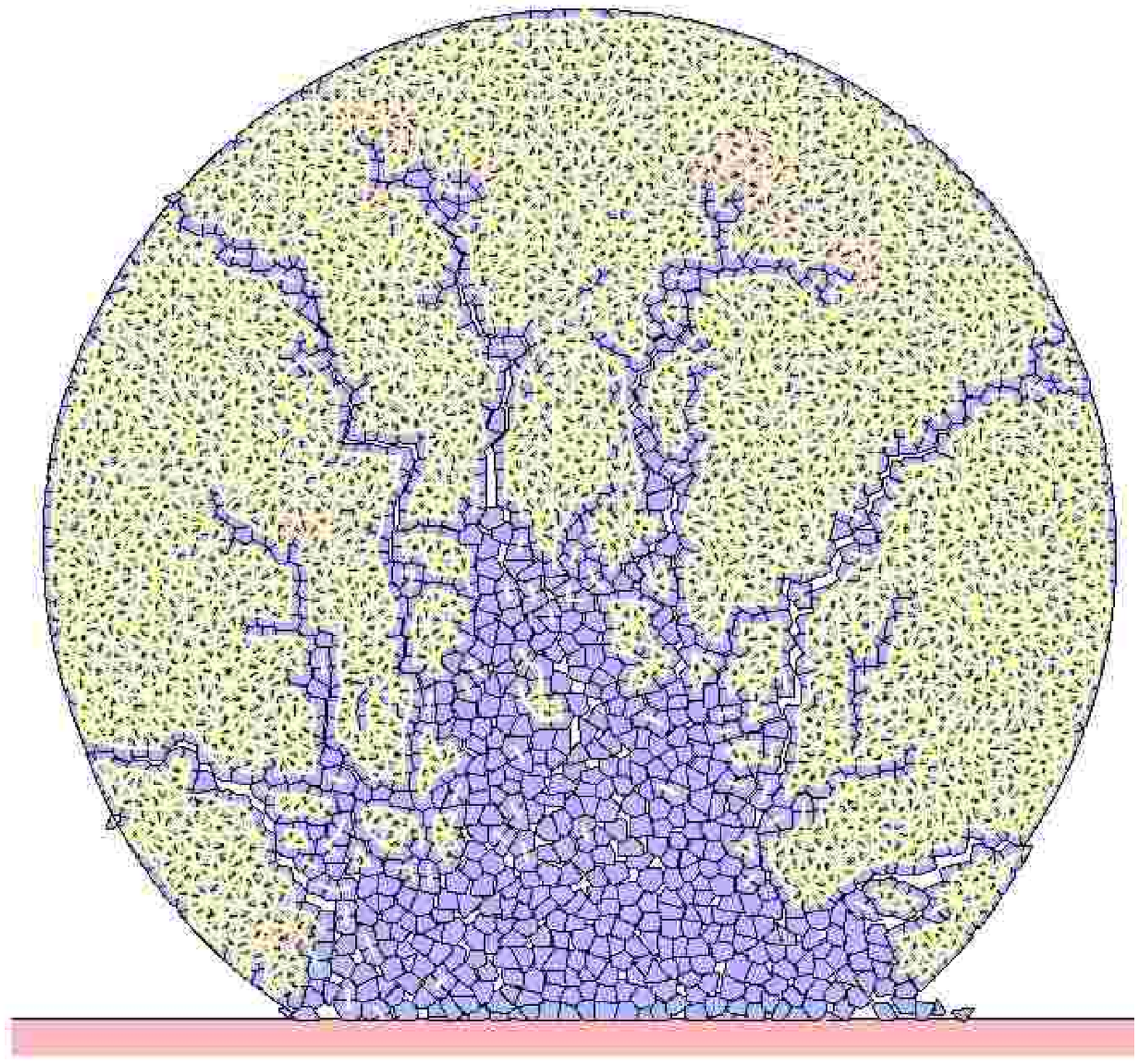}&
\includegraphics[width=0.4\textwidth]{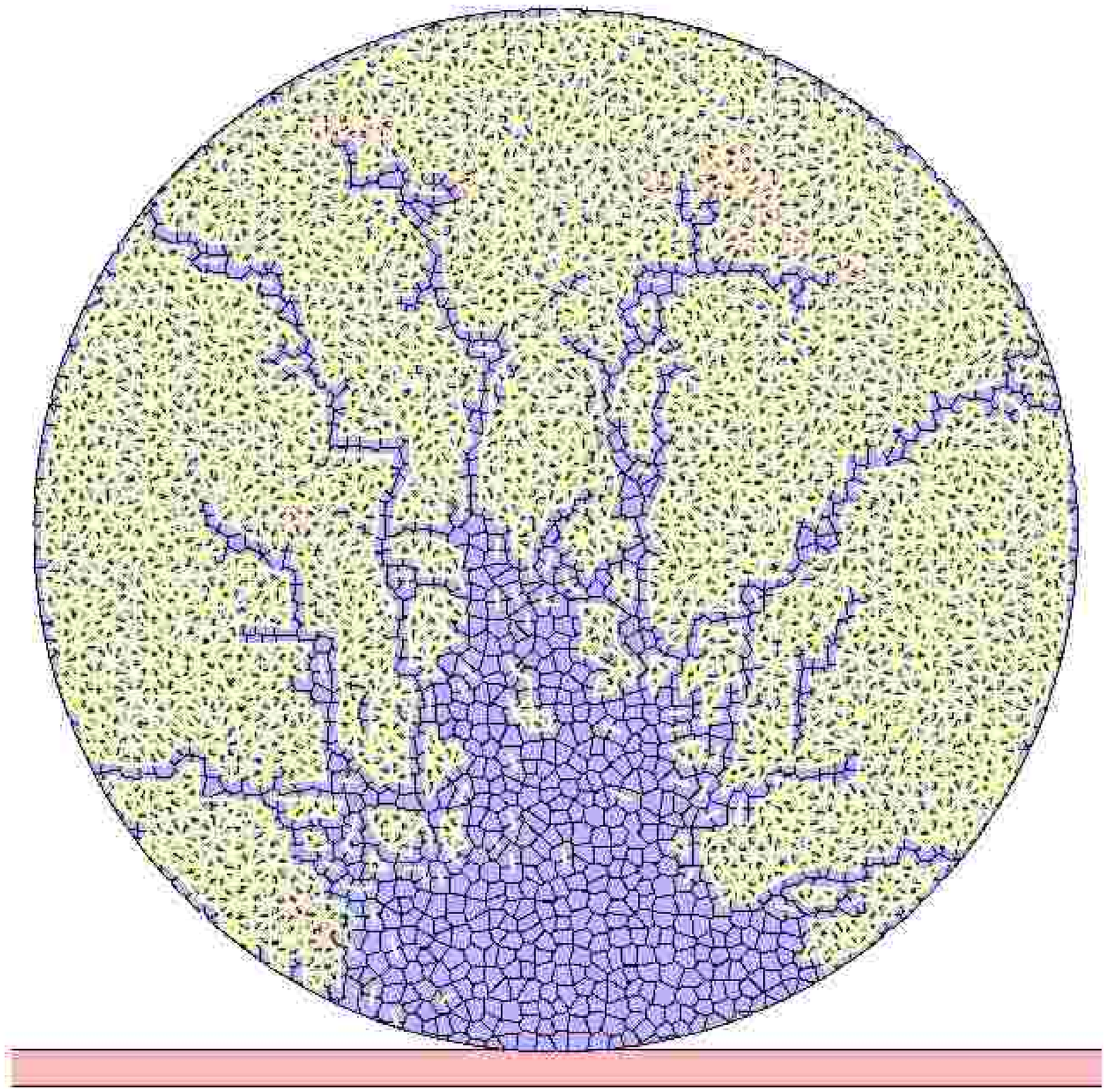}\\
(d)&(e)\\
\end{tabular}
\caption{\label{snapshots}
Snapshots of the fragmentation process in normal impact
 of a circular disc of radius 30 cm at $250$ cm/sec. 
(a) $t=0.00150$ sec, a high compressive wave generated at the impact
point causes the primary breakage.
(b) $t=0.00300$ sec, cracks start at the periphery of the conical
region whose base is the contact line between the disc and the plate.
(c) $t=0.00600$ sec, oblique cracks move outwards.
(d) $t=0.00996$ sec, the oblique cracks reach at the outer surface.
(e) The disc is reassembled at the end of the simulation to observe 
the final crack pattern. 
}
\end{figure}
In the present work we apply our model to explore the properties of
the fragmentation process of a circular disc when dropped on a 
frictionless hard plate at different angles. The particle moves with a
constant speed $v_0$ without the influence of gravity, which is obtained by  
supplying a constant velocity to all the polygons constituting the 
circular particle just before it touches the hard surface. The impact
angle $\theta$ defined as the angle of the vector of the impact
velocity to the horizontal, was varied between $90^{\circ}$ (normal
impact) and $45^{\rm o}$ (oblique impact).

In order to understand the break-up process of discs, we investigated
the crack pattern arising both in normal and oblique impacts. 
Fig.~\ref{snapshots} presents the time evolution of the crack pattern
of normal impact obtained by the simulation of a circular disc of
radius $30$cm at $250$cm/sec. When the disc strikes against the hard
plate, a high compressive wave is generated at the impact
point. Fracture starts from the region of contact point and propagates
through the disc. As the time passes, 
more and more bonds break at the impact region and  the
area of contact increases progressively. As a result of this primary
breakage a cone shaped (triangle shaped in two dimensions) damage area
is created whose base corresponds 
approximately to the area of contact of the specimen and the target 
(see Fig.~\ref{snapshots}b) and it is
more distinct at the end of the fragmentation process 
(see Fig.~\ref{snapshots}d). When the cone is driven into the specimen
a large number of cracks are generated starting from the region around
the cone (see Fig.~\ref{snapshots}b). This indicates that a high
stress concentration has developed around the conical damage region. 
Later on these  cracks run together to form few
oblique cracks (see Fig.~\ref{snapshots}c) directing  radially
outward. As crack propagation is very energy dissipative, when these
oblique cracks move outwards, the intensity of the compressive wave
gradually decreases and hence larger fragments appear 
opposite the impact point. 

\begin{figure}
\begin{tabular}{ccc}
\includegraphics[width=0.3\textwidth]{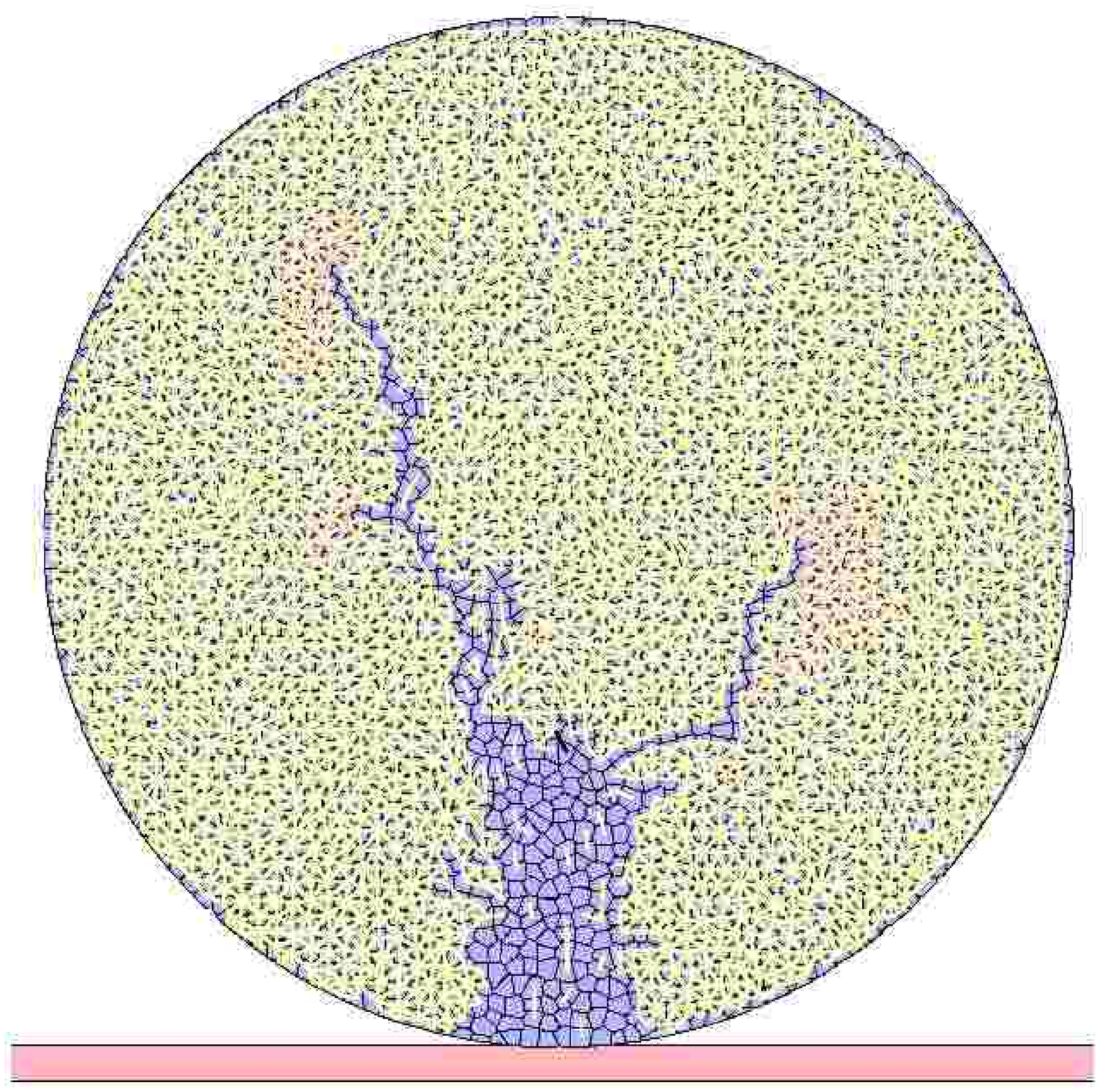}&
\includegraphics[width=0.3\textwidth]{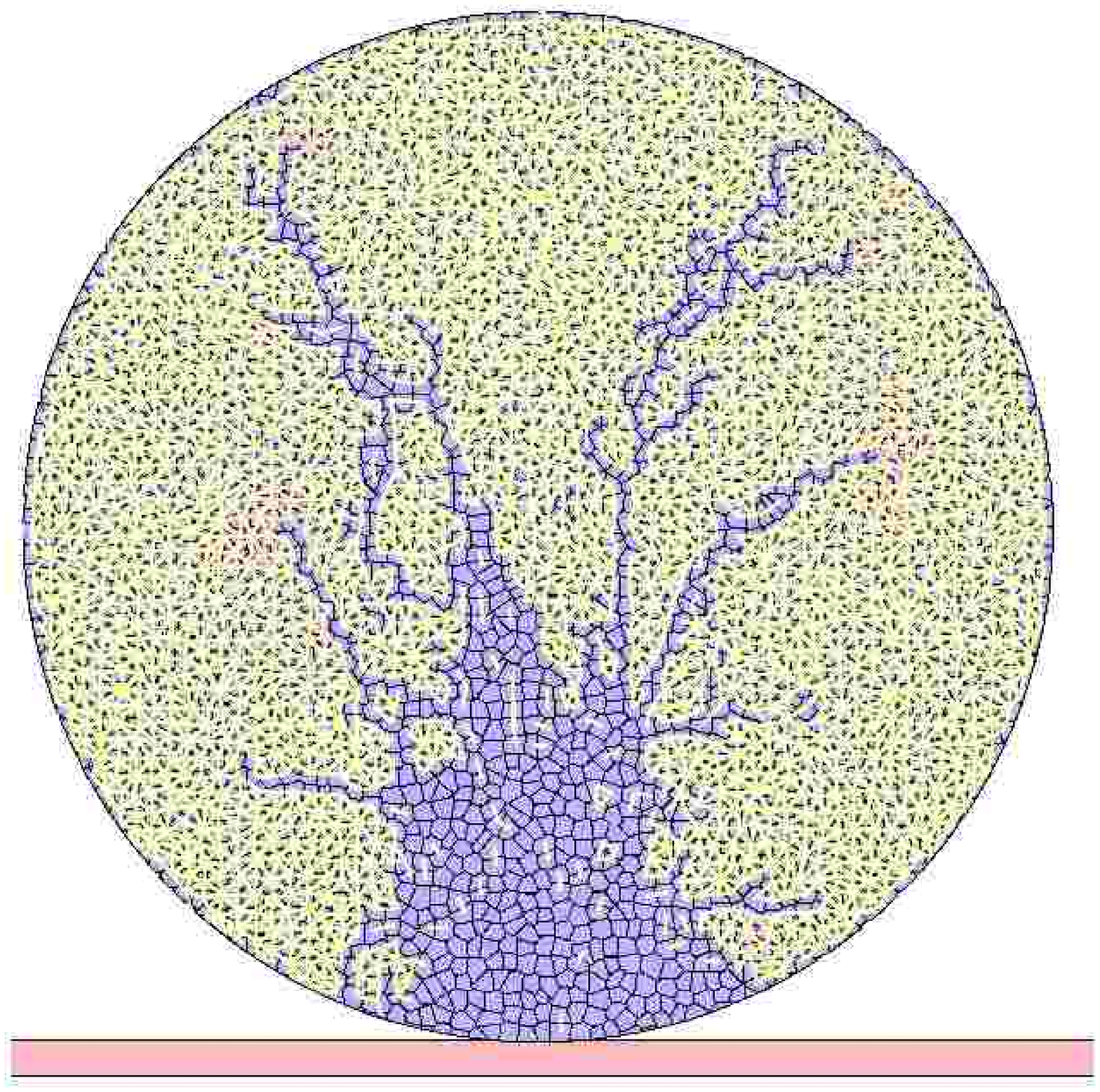}&
\includegraphics[width=0.3\textwidth]{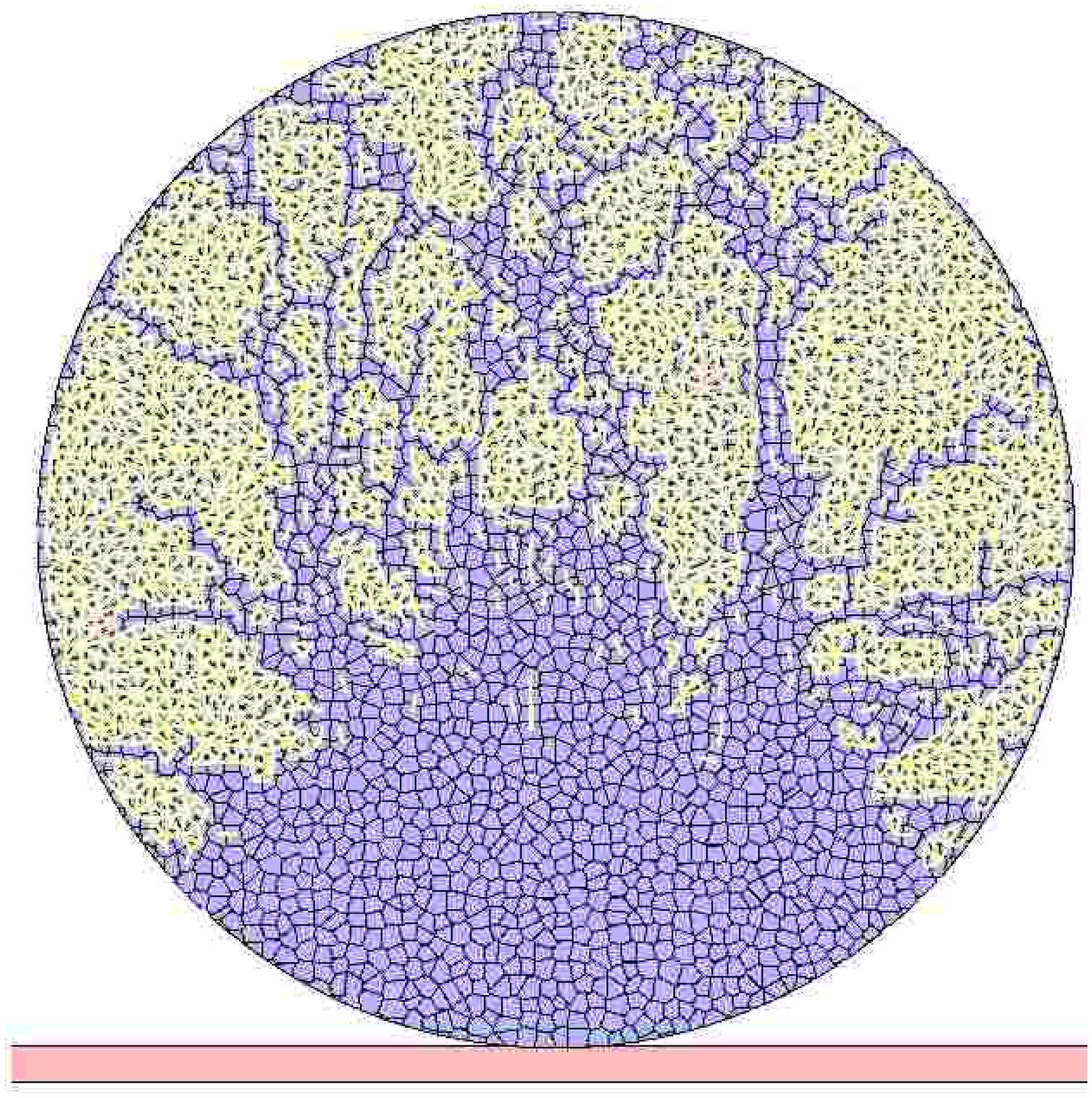}\\
(a)&(b)&(c)\\ 
\end{tabular}
\caption{\label{snapshot}
Final reassembled  states of the break-up process of a disc of radius
$30$ cm at different impact velocities dropped on a hard frictionless
plate.  
(a) $v_0 = 100$ cm/sec. The cone is not fully developed and only
few oblique cracks are present.
(b) $v_0 = 200$ cm/sec. More oblique cracks develop and
travel a greater distance.
(c) $v_0 = 600$ cm/sec. Both oblique cracks and secondary 
cracks are present. 
}
\end{figure}

To demonstrate the effect of  the impact velocity on the
break-up process, in Fig.\ \ref{snapshot} final states of the process
are shown obtained at different impact velocities, with the
fragments reassembled into the initial disc. 
At low velocities the compressive wave intensity generated at the
impact point is low, and hence, the cone could not develop
fully. Moreover, only a few oblique cracks are obtained, and they
do not reach  the opposite surface of the disc
(Fig.~\ref{snapshot}a). As the velocity increases, more oblique 
cracks develop and cover a greater distance (Fig.~\ref{snapshot}b) and
a considerable part of the initial kinetic energy goes into  the
motion of the residue resulting in rebound.
At the impact velocity where the oblique cracks reach the outer
surface of the disc opposite to the impact point, the break-up process
drastically changes: below this velocity mostly contact damage
occurs in the form of the damage cone and a relatively big residue
remains back. Above this velocity, however, the cracks spanning
the entire disc result in the break-up of the residue into
smaller pieces, see Fig.\ \ref{snapshots}d. Later on it will be shown
that the behavior of the system quantitatively changes at this
velocity, which we call critical velocity.
At impact velocities larger than the critical value, secondary cracks
are generated roughly perpendicular to the oblique 
cracks. Secondary cracks from neighboring oblique
cracks may  merge with each other as  can be seen in
Fig.~\ref{snapshot}c. Also at higher 
impact velocities, vertical cracks with a direction nearly 
perpendicular to the target plate are more prominent as the intensity of
stress concentration near the tip region of the cone is high as compared to
other parts. 

\begin{figure}
\begin{tabular}{cc}
\includegraphics[width=0.4\textwidth]{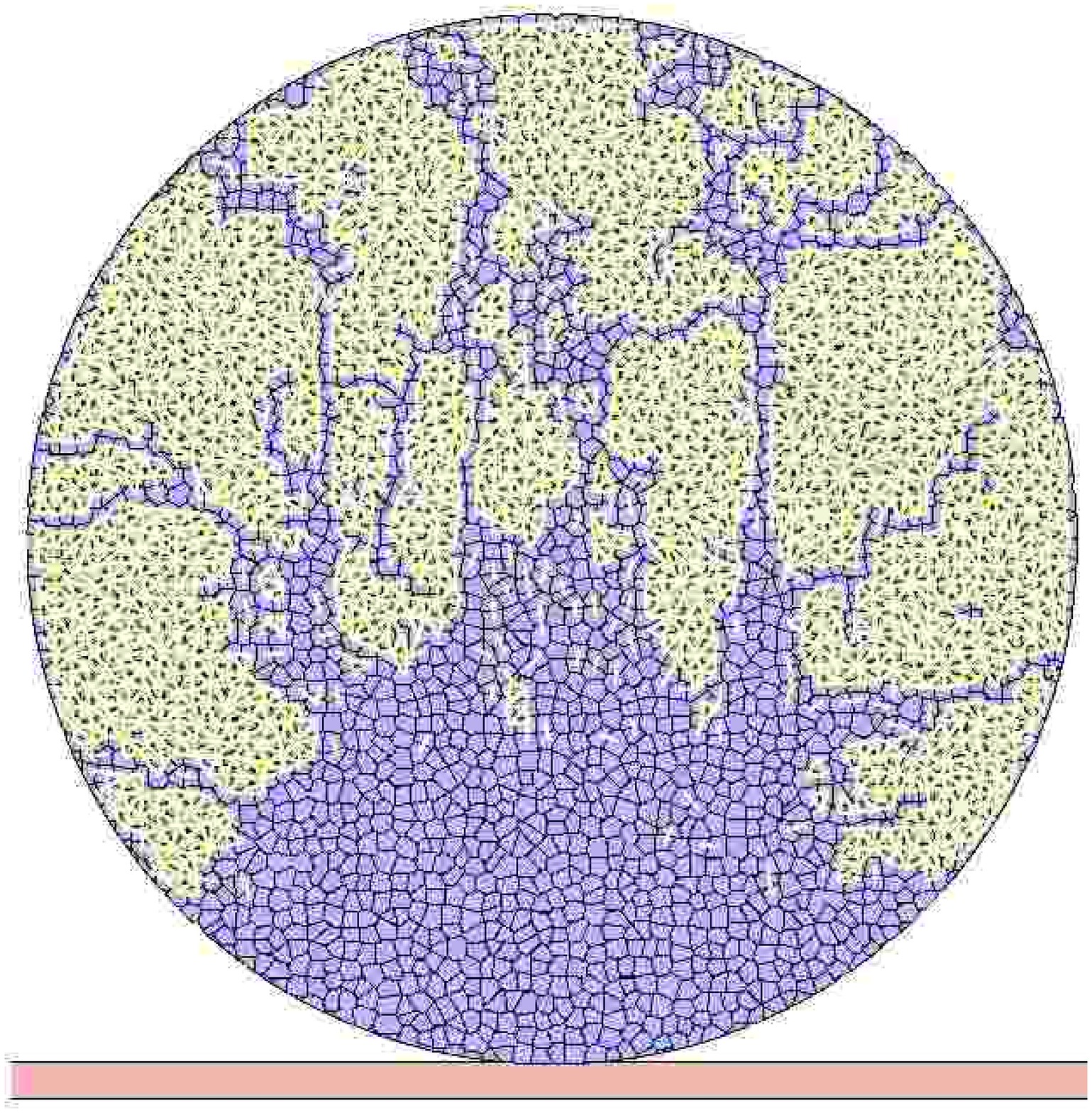}&
\includegraphics[width=0.4\textwidth]{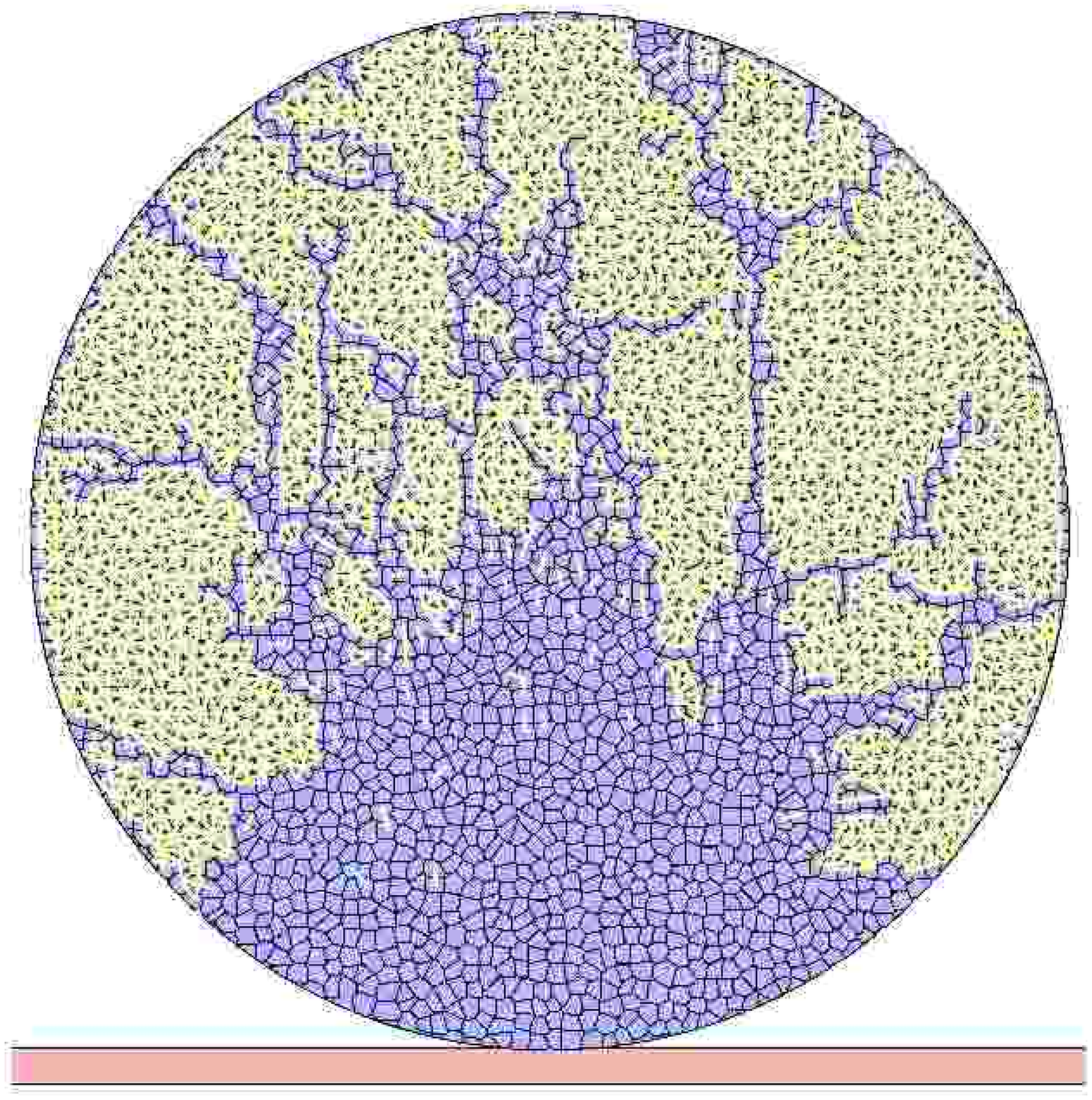}\\
(a)&(b)\\
\end{tabular}
\begin{tabular}{cc}
\includegraphics[width=0.4\textwidth]{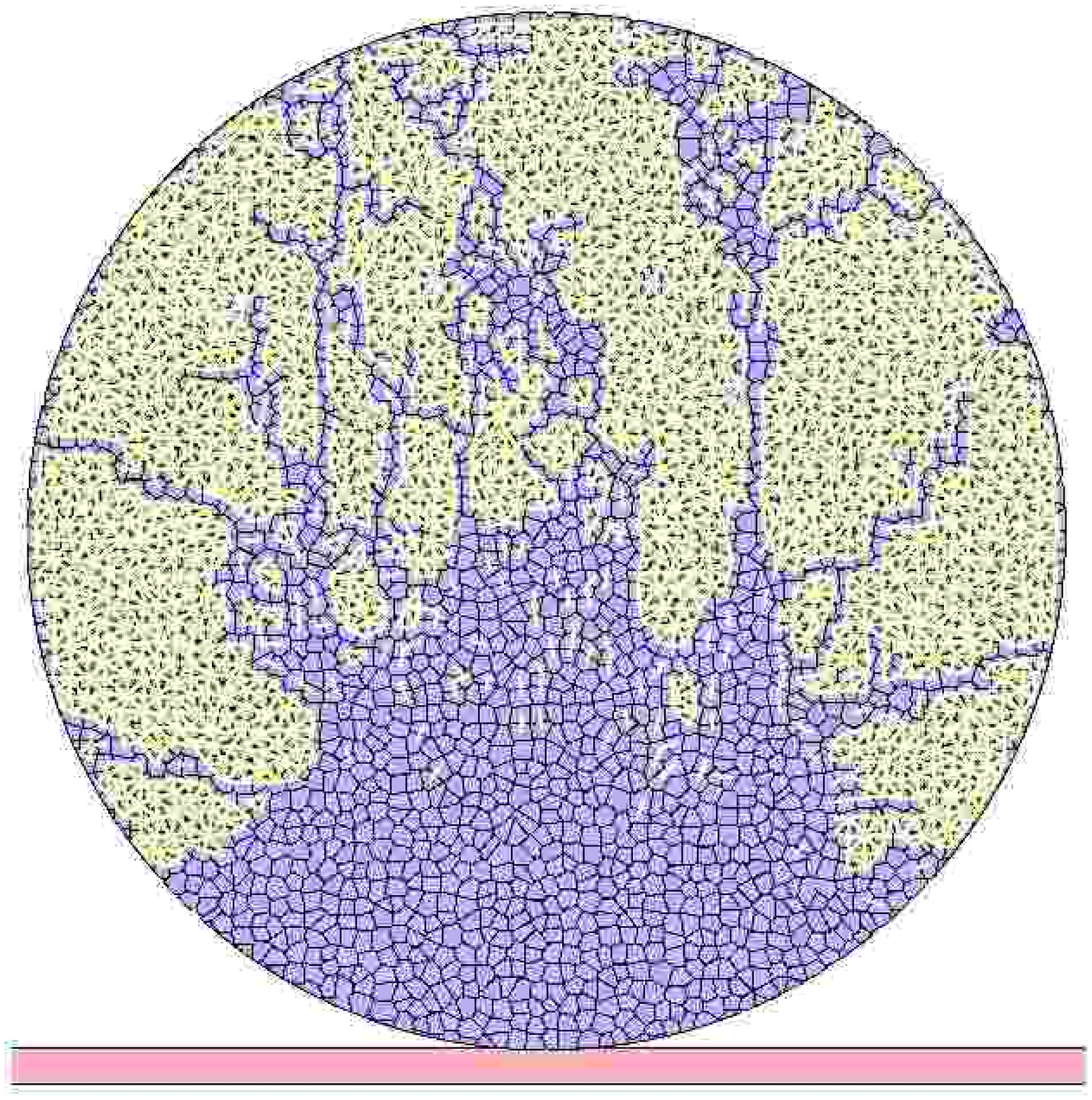}&
\includegraphics[width=0.4\textwidth]{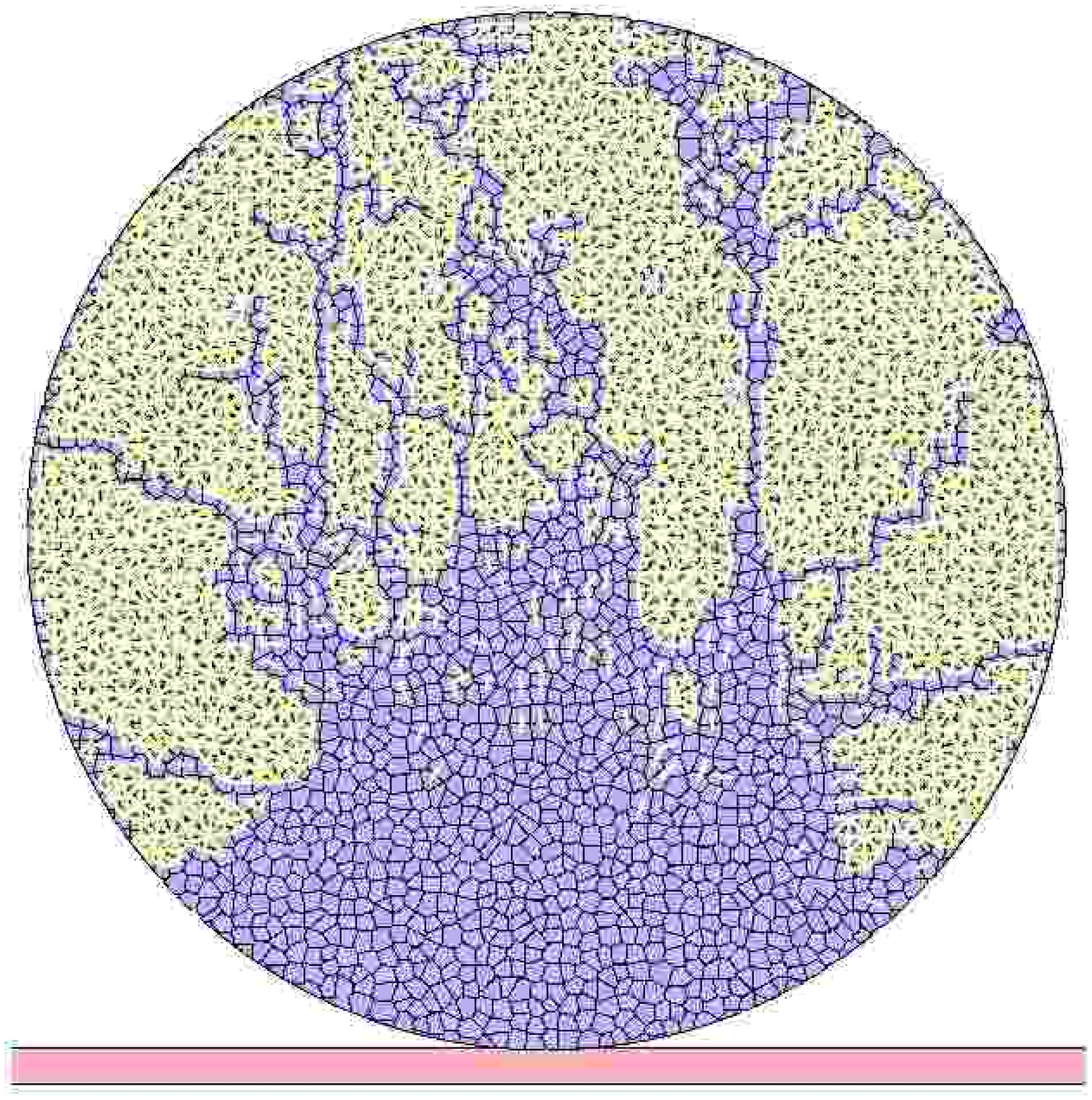}\\
(c)&(d)\\ 
\end{tabular}
\caption{\label{snap2}
The reassembled snapshots of normal and oblique impacts at different 
velocities keeping the normal component constant as $500$cm/sec. The crack
patterns are almost same in all cases.
(a) $\theta = 90^{\circ}$. (b) $\theta = 75^{\circ}$. 
(c) $\theta = 60^{\circ}$. (d) $\theta = 45^{\circ}$.
}
\end{figure}

Crack patterns obtained in the final state of oblique impacts at
impact angles $75^{\circ}, 60^{\circ}$ and $45^{\circ}$ are compared
in Fig.\ \ref{snap2}. It is important to emphasize that in our
calculations the friction between the target plate and the circular
disc is completely excluded. Under this condition, varying the impact
velocity while keeping its normal component constant, practically the
same crack pattern is obtained (see Fig.~\ref{snap2}). 
Thus, the crack propagation and orientation during the fragmentation
process solely depends on the normal component of the impact
velocity.
 
Comparing the crack pattern obtained in the simulations to the
experimental results \cite{fra_pattern}, we did not find any meridian
cracks as they are difficult to detect in two dimensions. The
pattern of oblique cracks and secondary cracks has a satisfactory
agreement with the experimental results. The simulations confirm that
oblique cracks which were observed in experimental investigations
\cite{fra_pattern} develop along the trajectories of maximum
compression planes. 
 
\section{Results}
\label{results}
Studying the evolution of the final crack patterns when changing the
impact velocity,  we have identified a critical
velocity $v_c$ which separates the regimes of different break-up
mechanisms. In the following, we analyze characteristic quantities of
the break-up process, and  show that there are substantial differences
between the two regimes.
\subsection{Size Distribution of Fragments}
\label{size}
\begin{figure}
\psfrag{aa}{damage}
\psfrag{bb}{fragmentation}
\begin{center}
\includegraphics[width=0.5\textwidth]{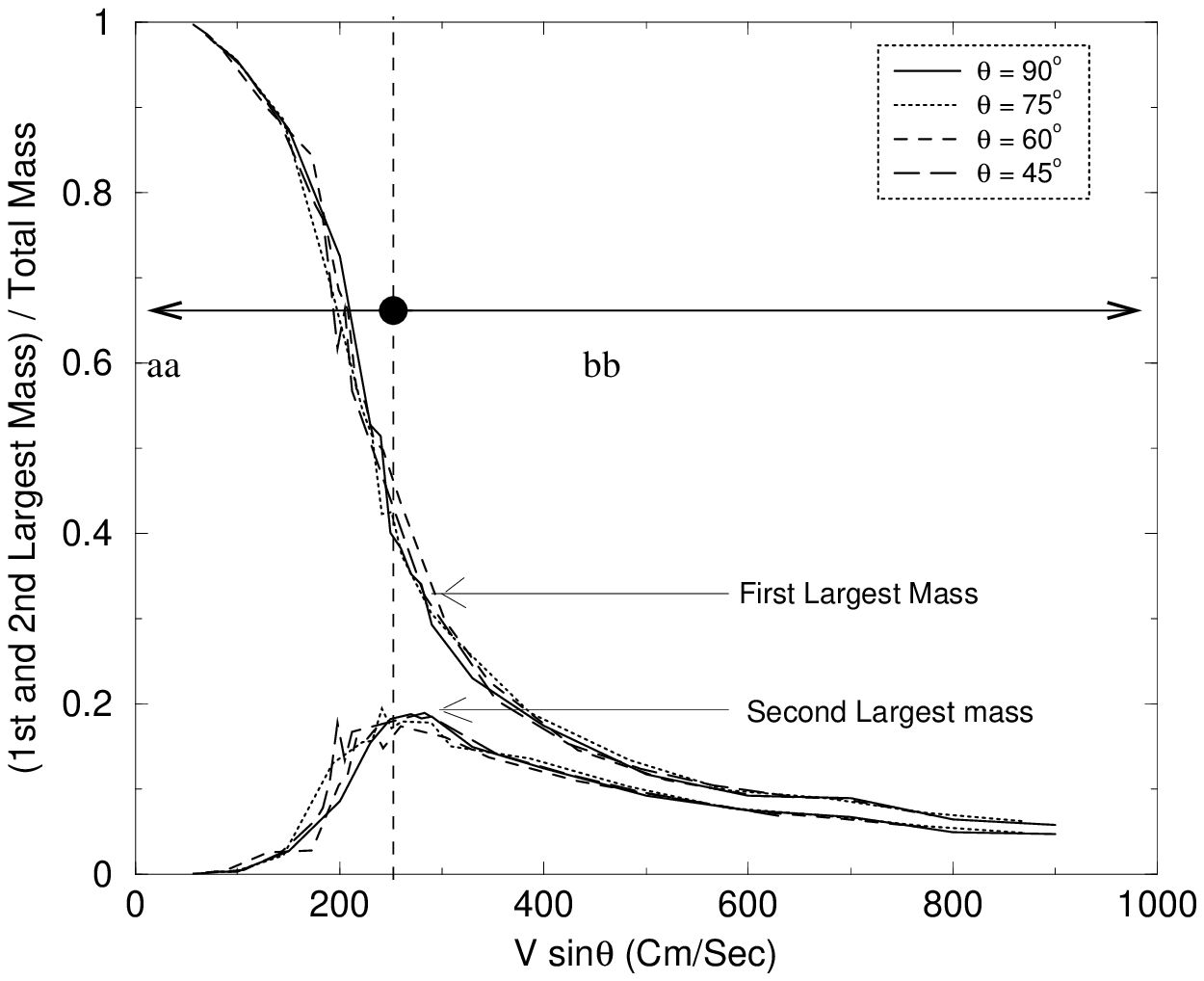}
\end{center}
\caption{\label{maxmas} The mass of the first and second largest
  fragment as a function of the normal component of the impact
  velocity $v_0\sin{\theta}$ at different impact angles $\theta$ and
  velocities $v_0$ for a system size $30$ cm. 
  The vertical dashed line indicates the critical point, furthermore,
  the damaged and fragmented regimes are also shown.
}
\end{figure}
Recently, it has been shown that the final outcome of a fragmentation
process can be classified into two states depending on the amount of
the imparted energy:  damaged and fragmented states with a sharp
transition in between. Detailed analysis revealed that the transition
between the two states occurs as a continuous phase transition
which also provides a possible explanation of the power law
mass distribution of fragments observed. 

To explore the nature of the critical velocity $v_c$ identified in the
previous section we investigated the evolution of the mass of the two
largest fragments when varying the angle $\theta$ and the velocity
$v_0$ of impact. Plotting the largest fragment mass as a function of
the normal component of the impact velocity 
for both normal and oblique impacts in Fig.\ \ref{maxmas}, all curves
fall over one another. This implies that in the absence of friction
between the plate and the disc, the size reduction achieved depends
both on $v_0$ and $\theta$ but in such a way that it depends on the
combination of the two variables $v_n = v_0 \sin{\theta}$. The curves
of the second largest mass exhibit the same data collapse when
plotting them as a function of $v_n$ further supporting the above
arguments. 

The functional form of the two largest fragment masses in Fig.\ \ref{maxmas}
shows the existence of two distinct regions.  
At low impact velocity,  breakage takes place  only at the impact point
and the largest fragment is nearly equal to the original mass. As the
velocity of impact increases, more small fragments are chipped off
from the impact point and cracks start around the
damaged conical region and move towards the outer surface of the
disc. At the critical velocity $v_c$ these propagating cracks reach 
the outer surface opposite to the impact point and the largest fragment
break into several big pieces.   The impact velocity where the
second largest mass attains its maximum value or where there is an
inflexion point in the largest fragment mass curve coincides with the critical velocity defined by
analyzing the cracking pattern in Figs.\
\ref{snapshots},\ref{snapshot}. In our case for normal impact 
of a system of $30$ cm radius the critical velocity turned out to be
$250$ cm/sec. 

The quality of the data collapse of the curves in Fig. \ref{maxmas} 
obtained at different
impact velocities $v_0$ and angles $\theta$ is excellent for the
largest fragment, however, there are large fluctuations of the value
of the second largest mass at impact velocities just below the
critical point. Above the critical point  all curves merge nicely
together. 

\begin{figure}
\begin{center}
\includegraphics[width=0.5\textwidth]{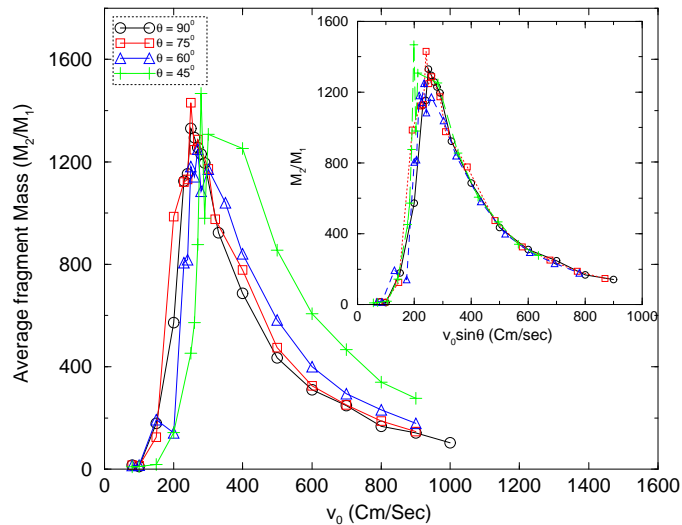}
\end{center}
\caption{\label{avgfra} The average fragment mass $\overline M$ as a 
function  of the impact velocity $v_0$. The inset shows the same
curves plotted as a function of the normal component of $v_0$.
 For oblique fragmentation when the velocity approaches 
the critical point always fluctuations arise, however beyond
the critical velocity all curves merge nicely.
}
\end{figure}
More information about the evolution of fragment sizes with varying
impact angle $\theta$ and velocity $v_0$ can be obtained by
studying the moments of fragment masses
\cite{instable,transition,univer,proj}.
The $k$th  moment $M_k$ of fragment masses is defined as
\begin{equation}
\label{eq:m_k}
M_k=\sum_i^{N_f} m_i^k-M_\mathrm{max}^k,
\end{equation}
where $N$ denotes the total number of fragments, $m_i$ is the mass of
fragment $i$ and $M_\mathrm{max}$
is the largest fragment mass. The definition Eq.\ (\ref{eq:m_k}) means
that the  $k$th power of the largest mass is
extracted from the sum of the $k$th 
power of the fragment masses. The average mass of fragments $\overline M$ can be defined
as the ratio of the second and first moments $\overline M \equiv
M_2/M_1$. In order to demonstrate the 
effect of rescaling the impact velocity, in the main panel of  Fig.\
\ref{avgfra} the average fragment mass $\overline M$ is presented as a
function of the impact velocity $v_0$ for the system size $R=30$cm
obtained at different impact angles $\theta$, and in the inset the
same curves are shown as a function of the normal component of
$v_0$. It can be seen that for each impact angle $\overline{M}$
has a peak which broadens and gets shifted
towards larger velocity values with decreasing impact angle. 
However, when plotting the same quantity as a function of the normal
components of the impact velocity $v_0\sin{\theta}$ all the curves
fall on  top of each other. Larger fluctuations arise below the
critical point which are more dominant for lower impact angles.
Note that the position of the maximum in the inset coincides with the
transition point determined in Fig.\ \ref{maxmas}.
\begin{figure}
\begin{center}
\includegraphics[width=0.5\textwidth]{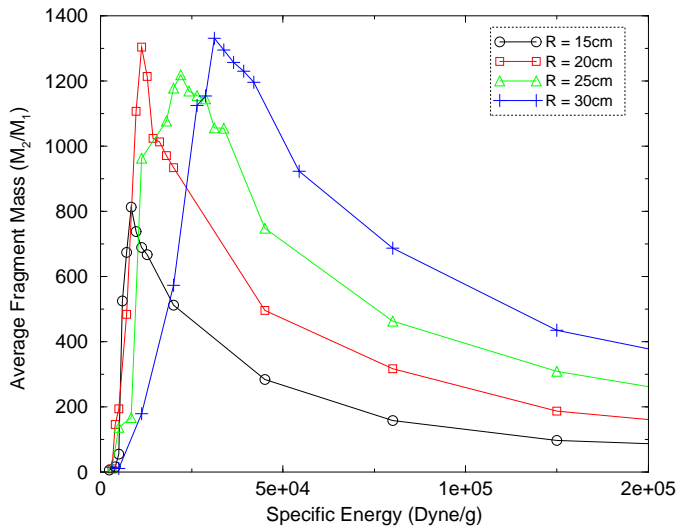}
\end{center}
\caption{\label{specific} Average fragment mass $\overline M$  as a 
function of specific energy $E_0/m_{tot}$. The critical point, {\it
  i.e.} the position of the maximum, depends on system size. 
}
\end{figure}

Studies of fragmentation of various types of brittle solids have revealed
that a larger amount of energy is required to achieve the same size
reduction on systems of larger size. However, in terms of the specific
energy, {\it i.e.} the energy imparted to the fragmenting system
divided by the total mass $E_0/m_{tot}$, all the characteristic
quantities show a universal behavior, especially the critical value of
the specific energy is independent of the system size
\cite{transition,proj,twodisc,granulate,poto,simu}.   
For impacting discs, however, the critical value of the specific
energy shows a clear dependence on the size of disc $R$ as it is
illustrated in Fig.\ \ref{specific}. The larger the disc, the higher
the energy density is required to break it into pieces. A
possible explanation is that for larger discs, a larger part of the
imparted energy goes into the motion of the fragments, lowering the
efficiency of break-up. 

\begin{figure}
\begin{center}
\includegraphics[width=0.5\textwidth]{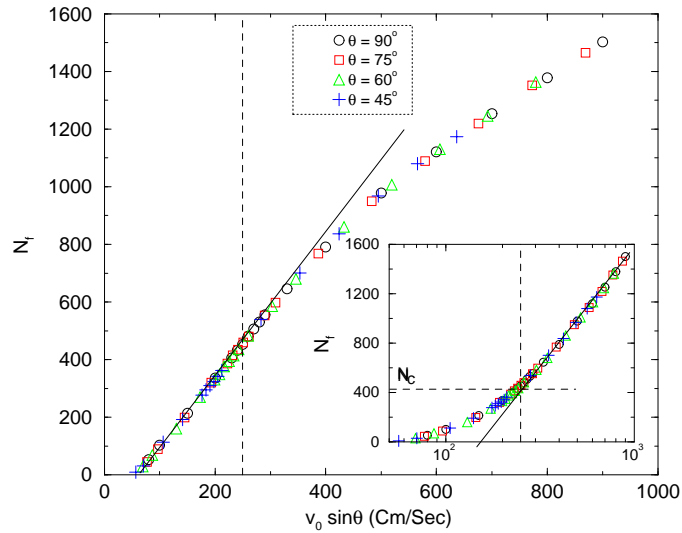}
\end{center}
\caption{\label{nofra} Number of fragments $N_f$ as a function of the normal 
component of the impact velocity.}
\end{figure}

The total number of fragments $N_f$ is also an important measure of
the degree of break-up in the impact process.
Fig~\ref{nofra} shows that the number of fragments $N_f$ is uniquely
determined by the normal component of the impact velocity $v_n$, {\it
  i.e.} the curves obtained at different impact angles present a
perfect collapse when plotting them as a function of $v_n$. 
It can be seen in the figure that the number of fragments is a
monotonically increasing function of the velocity, however, the
functional form of $N_f$ seems to be different on the two sides of the
critical point, {\it i.e.} up to the critical point the curves show
clearly a straight line, whereas, above the critical point all curves
are slightly bent towards down as the efficiency of the fragmentation 
process decreases. Replotting the results using logarithmic scale on
the horizontal axis, however, a straight line is obtained above the
critical point (see the inset of Fig.\ \ref{nofra}), which implies
that $N_f$ has the form
\begin{eqnarray}
N_f = a\cdot \ln{\frac{v_n}{v_{nc}}} + N_c, \ \ \mbox{for} \ \ v_n > v_{nc},
\end{eqnarray}
where $N_c$ denotes the number of fragments at the critical point and
$a$ is the slope of the straight line in the inset of Fig.\
\ref{nofra}. 

\begin{figure}
\begin{center}
\includegraphics[width=0.5\textwidth]{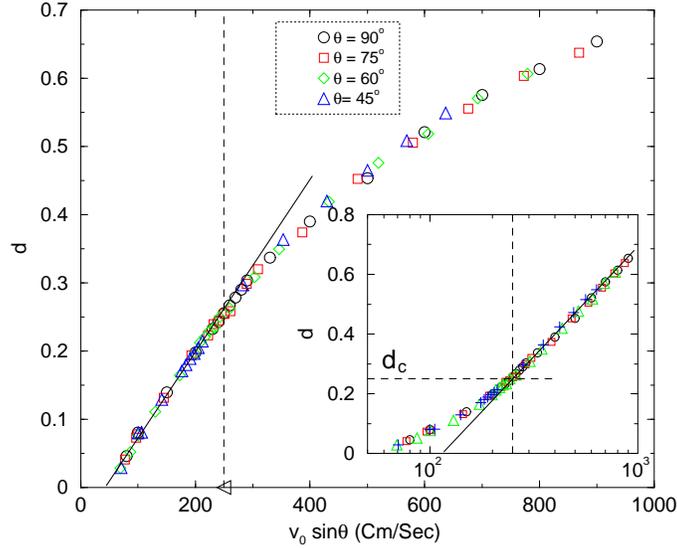}
\end{center}
\caption{\label{damage_ratio} Dependency of damage ratio $d$, the number of
broken beams $N_b$ to the total number of beams $N_o$, on the normal 
component of the impact velocity.}
\end{figure}

The amount of damage occurring during the break-up process can be
quantified by the so-called damage ratio $d$ proposed by   
Thornton {\it et.\ al} \cite{agglomerate}. $d$ is defined
as the ratio of the number of broken contacts $N_b$ to the total 
number of contacts $N_c$ existing initially inside the disc. The
damage ratio $d$ depends both on the impact angle $\theta$ and the
impact velocity $v_0$, {\it i.e.}, increasing $v_0$ at a fixed value
of $\theta$ results in an increase of $d$, furthermore, increasing the
impact angle $\theta$ at a given value of $v_0$ the damage ratio also
increases. However, when plotting $d$ as a function of the normal
velocity $v_n$ in Fig.~\ref{damage_ratio} the curves obtained at different
impact angles collapse on top of each other, which implies that
$d$ solely depends on $v_n$. Similarly to the number of fragments, $d$
is also a monotonically increasing function of $v_n$, however, its
functional form changes at 
the critical point. It is observed that below the critical point $d$
is a linear function of $v_n$, while above the critical point the curve is 
non-linear, slightly bending down. On a semilogarithmic plot
again a straight line arises which implies the functional form
\begin{equation}
d = b\cdot \ln{\left(\frac{v_n}{v_{cn}}\right)}+d_c, \ \ \mbox{for} \
\ v_n > v_c
\end{equation}
where $d_c$ is the value of $d$ at critical point and $b$ is the slope
of the fitted straight line in Fig.\ \ref{damage_ratio}. A somewhat
similar functional form of $d$ has also been pointed out by Thornton 
\cite{agglomerate,imp_agglo} in impact of discs and spherical objects
with a hard plate. 
\subsection{Plate Force}
\begin{figure}
\begin{tabular}{cc}
\includegraphics[width=0.49\textwidth]{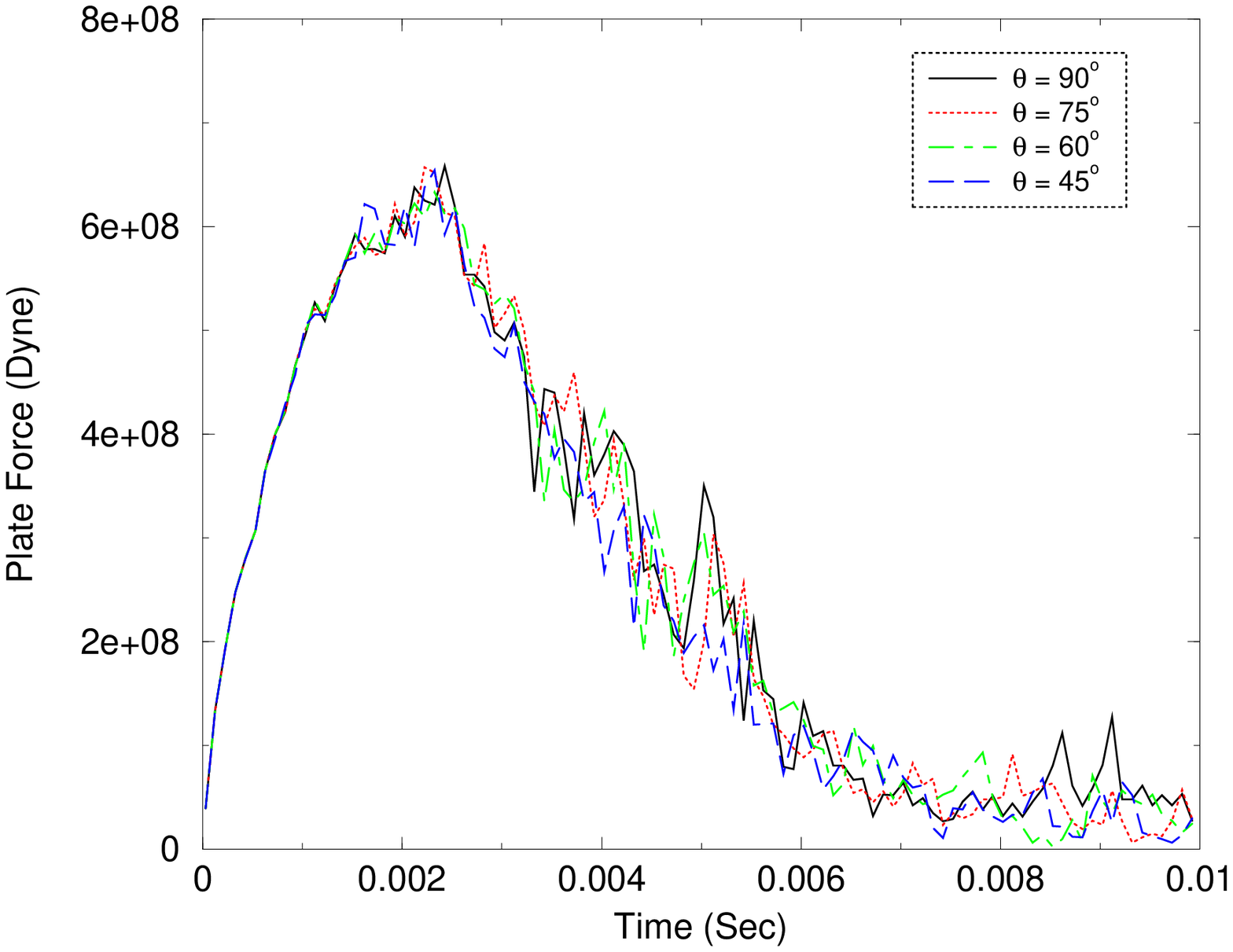}
\includegraphics[width=0.49\textwidth]{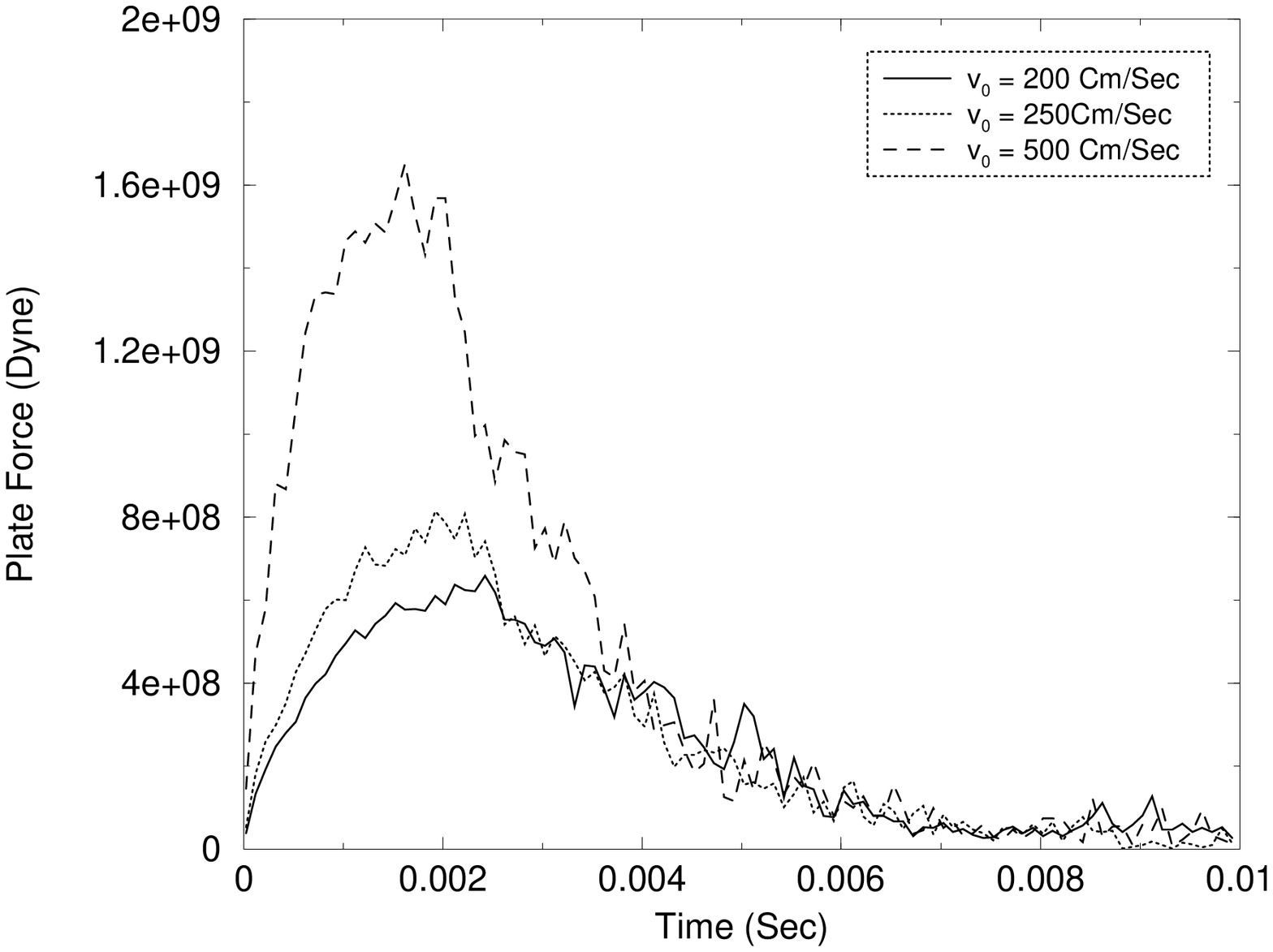}\\
(a)&(b)\\
\end{tabular}
\begin{tabular}{cc}
\includegraphics[width=0.49\textwidth]{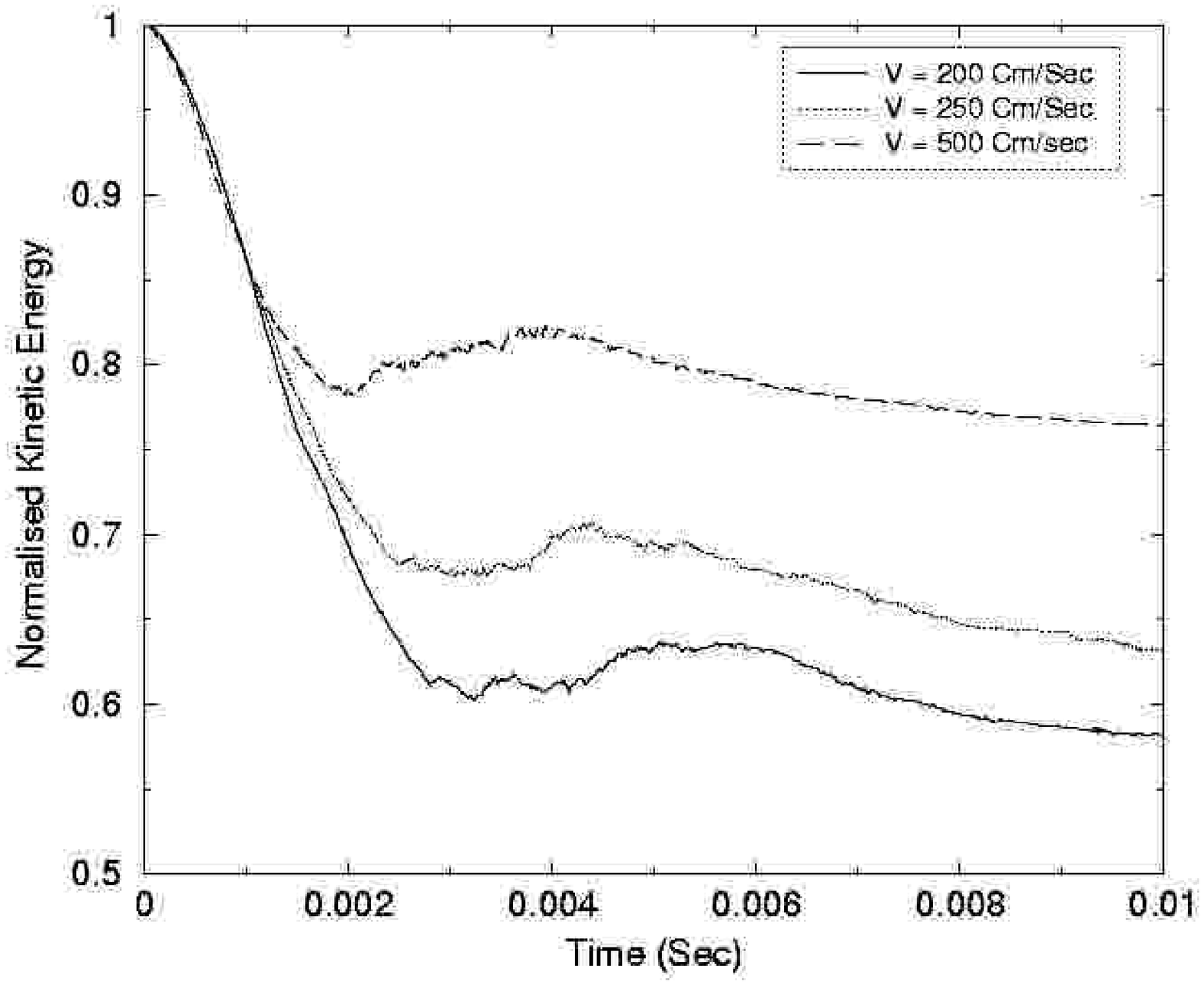}&
\includegraphics[width=0.40\textwidth]{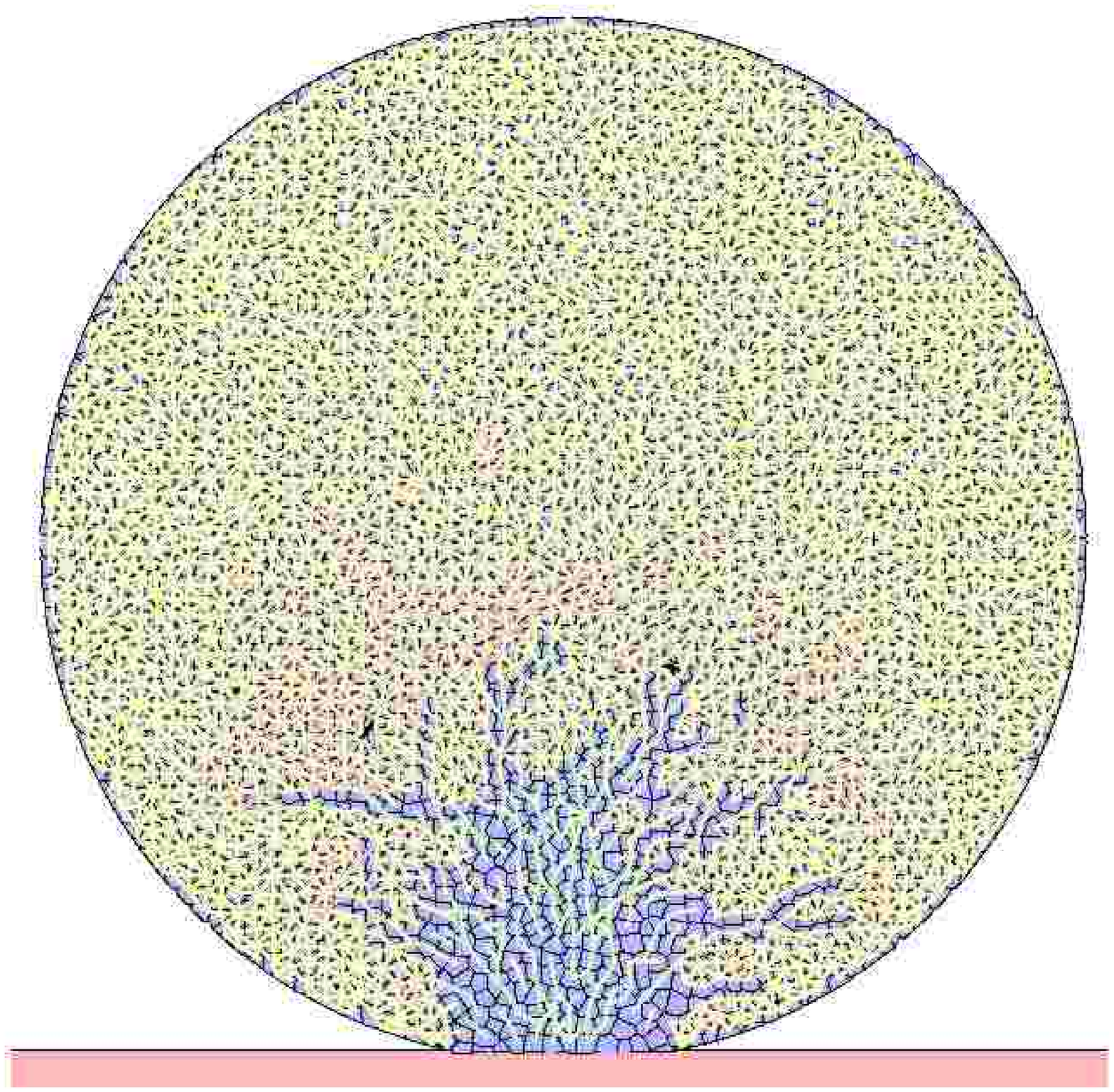}\\
(c)&(d)\\
\end{tabular}
\caption{\label{wforce}
(a) Keeping the normal component of impact
velocity constant at different impact angle, the force exerted by the plate is
almost the same. (b) Force exerted by the plate at various velocities during
normal impact. (c) The kinetic energy normalized by the 
total kinetic energy as a function of time.
(d) The state of the disc of size $30$cm after $t=0.00210$sec 
which corresponds to the maximum plate force and minimum kinetic energy
at the impact velocity of $250$cm/sec.
}
\end{figure}
An alternative way of showing the effect of the impact angle is by analyzing
the force exerted by the plate on the disc. In Fig.~\ref{wforce}(a) and (b)
typical time series of the force between the target plate and the disc
are presented.
If the  normal component of the impact velocity $v_n =
v_0\sin{\theta}$ is kept constant while changing $\theta$ and $v_0$,
the force practically remains constant except for fluctuations, see
Fig.~\ref{wforce}(a). 
 It shows clearly that the force exerted by the target plate only 
depends on the normal component of impact velocity providing
further support for the above findings in consistency with refs. 
\cite{agglomerate,bk,imp_agglo,imp_ang}.  

To take a clear view of the nature
of the plate force, we have plotted the plate force as a function of
time at various velocities (see Fig.~\ref{wforce}(b)).
 In general, as the impact
velocity is increased, the maximum plate force increases, the duration of 
the impact decreases. The maximum force exerted by the plate occurs when
the kinetic energy has a minimum (see Fig.~\ref{wforce}(c)). The maximum
plate force or minimum kinetic energy corresponds to the state of the
fragmented disc where most of the bonds break near the contact region
and cracks start to propagate radially outwards from the conical
damage region (see Fig.~\ref{wforce}(d). 
Since damage, {\it i.e.} bond breaking dissipates energy, the final
kinetic energy is significantly less than the initial kinetic energy. 
Increasing the impact velocity gives rise to an increase in the final
kinetic energy  and decrease of the duration of plate-disc contact.  
Moreover, the initial kinetic energy remaining at the end of the impact first
 decreases 
with impact velocity until the velocity is sufficient to produce multiple
fracture and then increases due to increase in kinetic energy of the broken
fragments. Clearly, there are two stages to the bond breaking process during 
impact. Initially bonds are broken primarily as a result of the high
compressive 
shock wave adjacent to the impact site, which occurs during the period when
 the plate force is increasing. This is followed by further bond breakage 
due to crack propagation radially outwards, starting around the conical
 damage region where high stress concentration occurs, while the plate force 
decreases.
\subsection{Mass Distribution of Fragments}
\begin{figure}
\begin{center}
\includegraphics[width=0.45\textwidth]{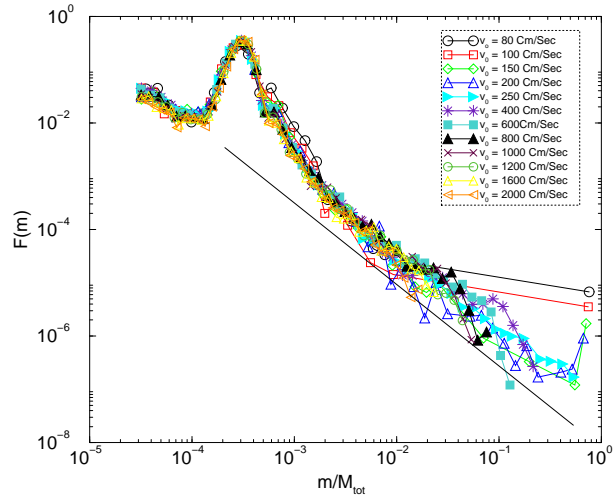}
\end{center}
\caption{\label{mass} The fragment mass histograms of normal impact 
for the system size $R = 30$cm with varying initial impact velocity. The
straight line shows the power law fitted to the curve at the critical 
velocity $v_0 = 250$cm/sec with an exponent $\tau = 1.835$.}
\end{figure}

The mass (or size) distribution of fragments is one of the most
important characteristic quantities of the disc impact which has also a
high practical relevance.
The fragment mass histograms $F(m)$ of normal impact are presented in the
Fig~\ref{mass}, for the system size $30$cm at varying impact
velocity $v_0$. In order to resolve the shape of the distribution 
over a wide range of mass values, logarithmic binning was used {\it
  i.e.,} the binning is equidistant on logarithmic scale. 

It can be observed that the histograms have a maximum at small
fragment sizes due to the existence of single unbreakable
polygons. The shape of the peak of the small fragments is determined
by the mass distribution of single Voronoi polygons obtained by the
tessellation procedure.
At low velocities, much below the critical point, the distributions are
discontinuous: for small fragment masses the distributions are smoothly
decreasing functions, while for large fragments $F(m)$ has a peak
indicating the presence of large unbroken pieces. In between however,
the medium sized fragments are missing. As the impact velocity
increases, the large pieces break up into smaller ones, the
peak of the large fragments on the right hand side gradually
disappears and the entire mass distribution becomes continuous. It is
interesting to note that the peak of large fragments disappears
completely at the critical point where the cracks starting from 
the damaged conical region reach at the outer surface of the disc, breaking
the disc into several smaller pieces. As a result, $F(m)$ takes a
power law form at the critical point 
\begin{eqnarray}
F(m) \sim m^{-\tau}. 
\end{eqnarray}
The exponent of
the power law $\tau$ fitted  
to the curve at the critical velocity $v_0 = 250$cm/sec is $\tau =
1.835$. For  
oblique impact the value of the exponent is nearly the same as in 
normal impact within a precision of $\pm 0.05$. Simulations with
different system sizes such as $R=25$cm, $R=20$cm and $R=15$cm proved
that the exponent $\tau$ is also independent of $R$.

Increasing the impact velocity above the critical point, the power
law regime of the mass distribution remains unchanged, however, the
largest fragment size decreased and the shape of the curve attains
an exponential form for large fragments. In the limiting case of very
high impact velocities the disc suffers complete disintegration into
small pieces. In this shattered phase $F(m)$ gradually transforms to
an exponential form. However, the shattered phase is slowly approached when
increasing the impact velocity $v_0$ since the damage ratio and the
number of fragments have a logarithmic dependence on $v_0$.
The results are in good quantitative agreement with 
the experimental findings on the fragmentation of plate-like objects
\cite{glassplate,platelike,composite,ice}.

\subsection{Scaling of the Velocity Distribution}
In applications like mineral processing, a fragment, formed with a
certain velocity, can undergo secondary break-up due to collisions with
other fragments or with the walls of the container. To get an estimate
about the importance of this secondary fragmentation, it is essential
to study the fragment velocities. We investigated the velocity
distribution 
of fragments and its dependence on the macroscopic variables of the
system like the impact velocity $v_0$ and radius of the disc $R$.
\begin{figure}
\begin{tabular}{cc}
\includegraphics[width=0.49\textwidth]{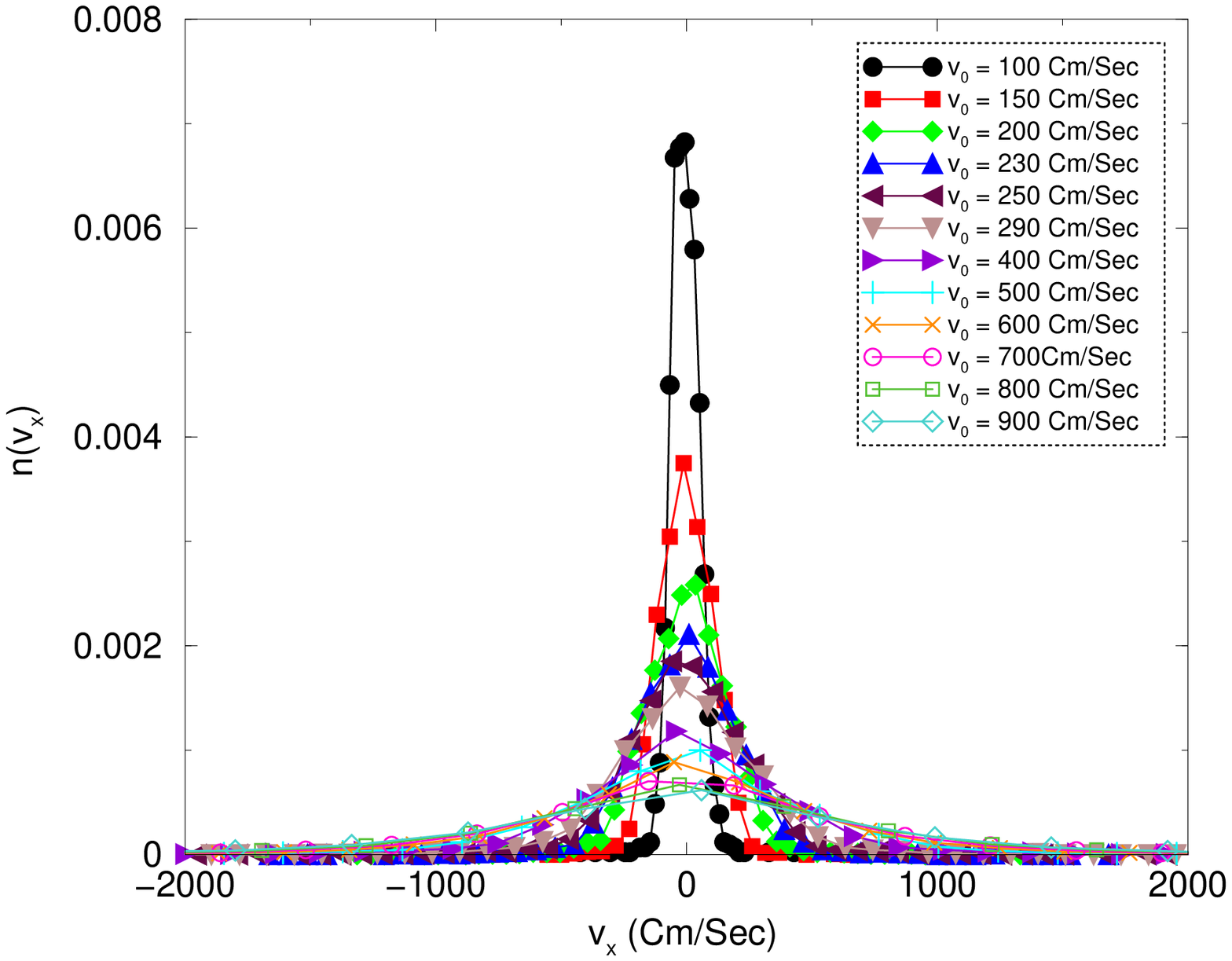}&
\includegraphics[width=0.49\textwidth]{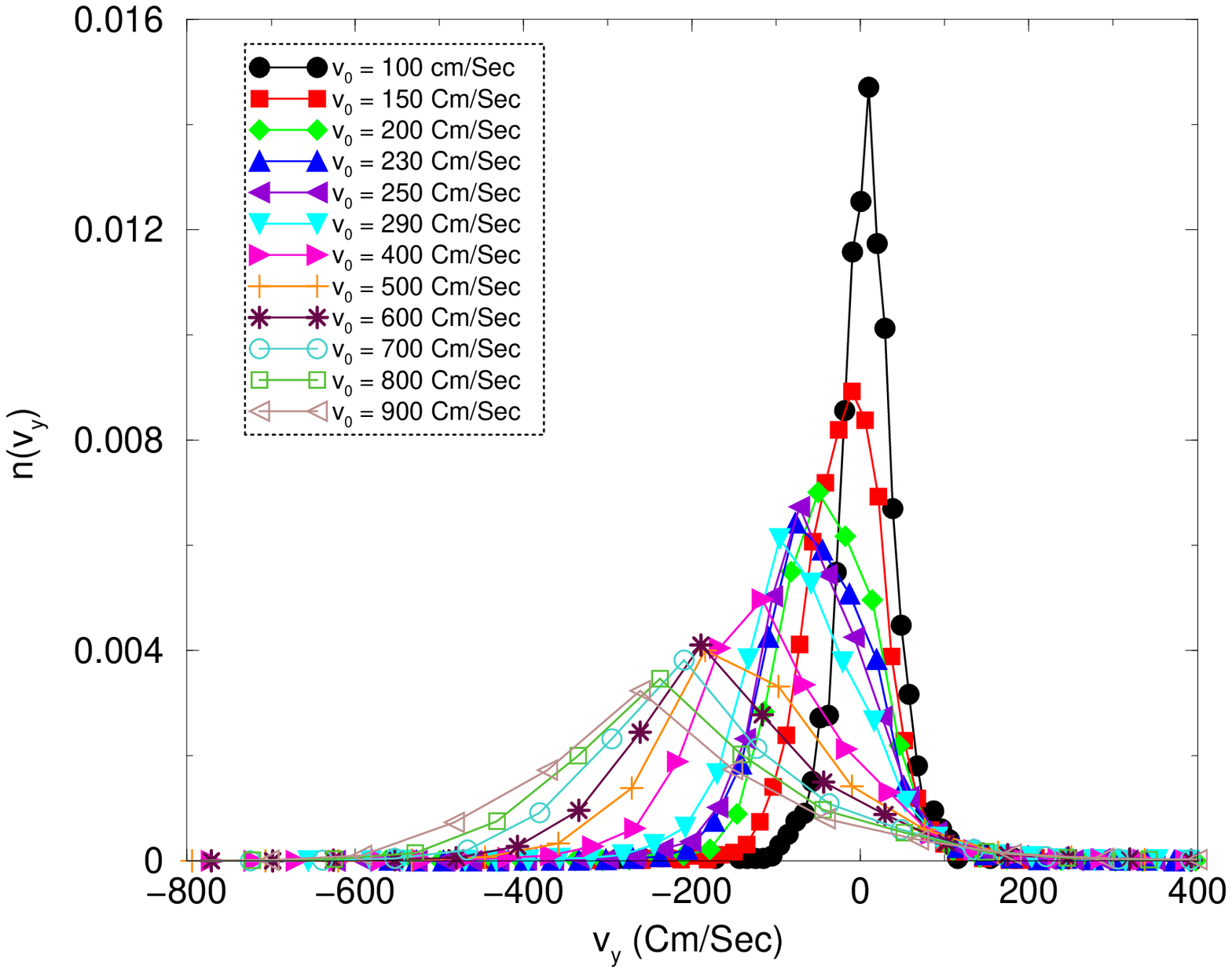}\\
(a)&(b)\\
\end{tabular}
\caption{\label{veldistv} The distribution of the $x$ and $y$ components 
of the velocity of fragments with fixed system size R=$30$ cm varying 
the initial impact velocity $v_0$.}
\end{figure}
\begin{figure}
\begin{tabular}{cc}
\includegraphics[width=0.49\textwidth]{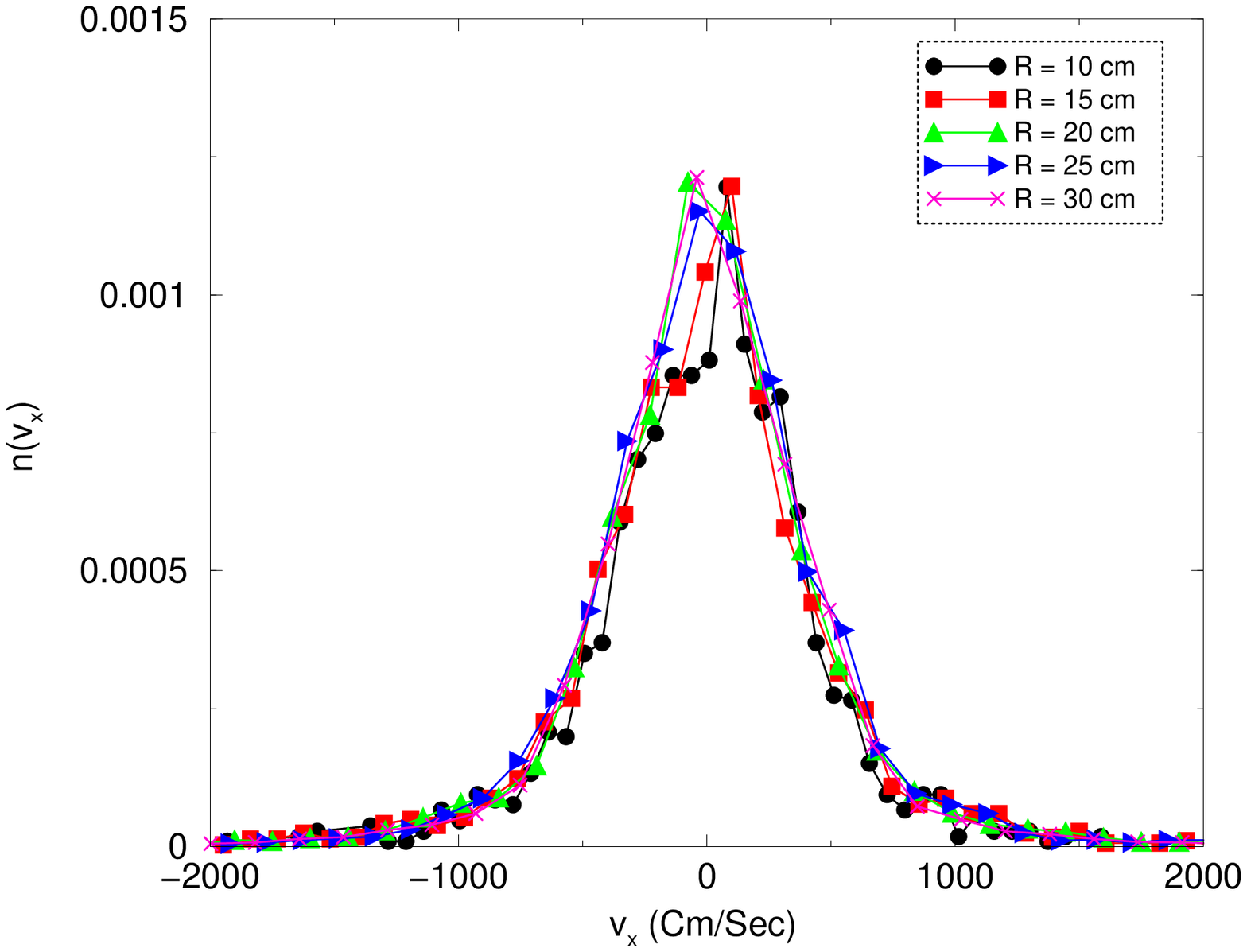}&
\includegraphics[width=0.49\textwidth]{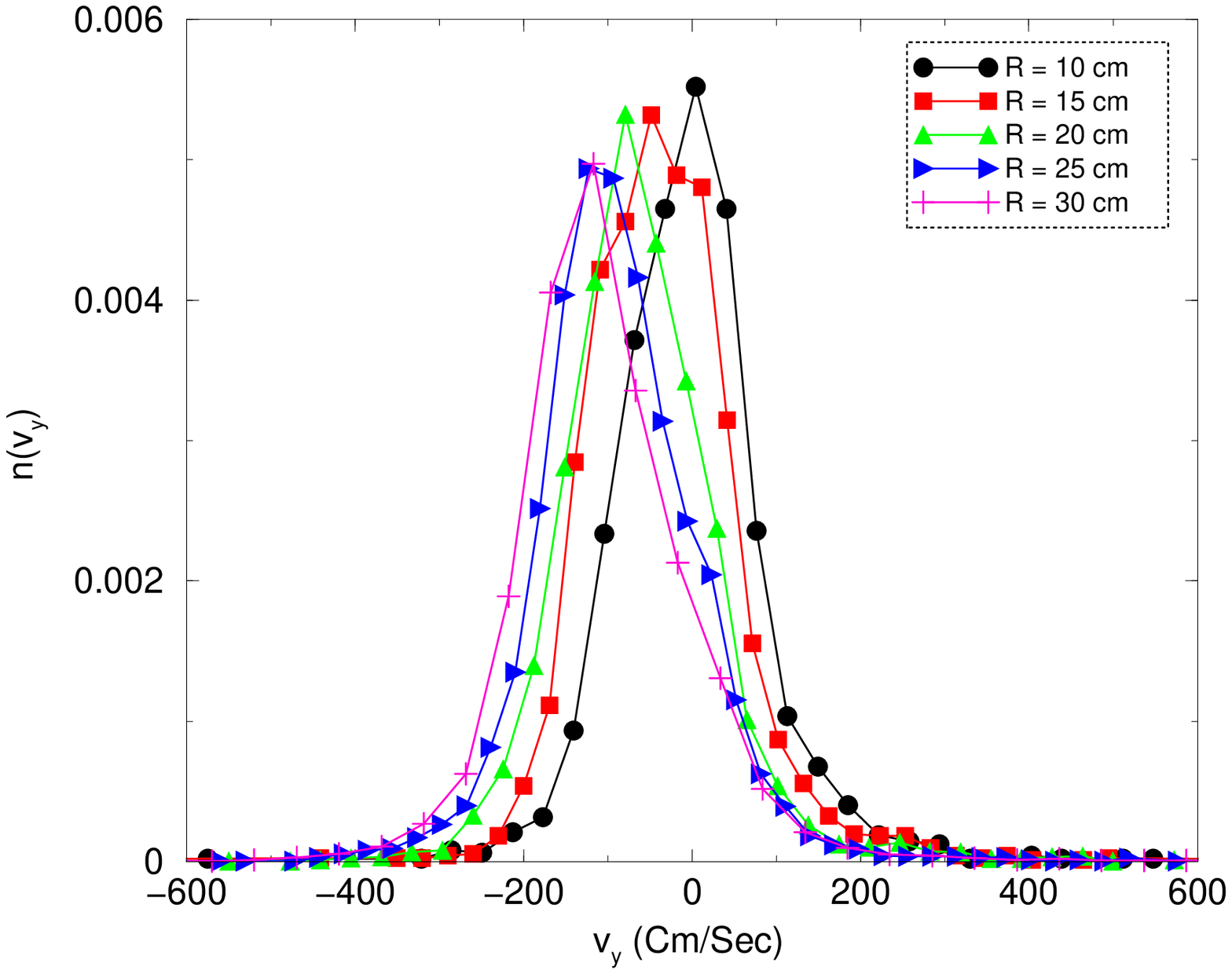}\\
(a)&(b)\\
\end{tabular}
\caption{\label{veldistr} The distribution of the $x$ and $y$ components 
of the velocity of fragments with fixed impact velocity $v_0 = 400$
cm/sec varying the radius $R$ of the particle.}
\end{figure}
To determine the velocity distribution of fragments and explore its
dependence on  $v_0$ and  
$R$, we analyzed the data in two ways. First we fixed the disc radius
$R$ and varied the impact velocity $v_0$, then fixed $v_0$ while
varying $R$. In both cases the calculations are restricted to normal
impact ($\theta = 90^{\circ}$) and the distributions of the velocity
components $n(v_x)$, $n(v_y)$ of the center of mass of the fragments
are evaluated. In Figs.~\ref{veldistv},\ref{veldistr} we 
present the results for fixed radius $R = 30$ varying the initial
velocity and for fixed $v_0 = 400$ cm/sec varying the radius of
the particles, respectively.
In Fig.~\ref{veldistv}(a) one can observe that the distribution of the
$x$ component of the fragment velocities $n(v_x)$ is symmetric about
$v_x=0$ as expected from the symmetry of the initial conditions. The
zero mean value is a consequence of momentum conservation. As the
impact velocity increases, the distribution broadens.  
The distribution $n(v_y)$ of the $y$ components also broadens with
increasing impact velocity but also shifts towards the negative
$y$ direction. This is obvious as the direction of the impact velocity is 
in the negative direction of $y$ axis and the total linear momentum
increases with the impact velocity. However, in the $y$ direction
fragments are slower, {\it i.e.,} the values of $v_y$  are much
smaller than those of $v_x$. Note that there is  a small fraction of
the debris which has velocity larger than $v_0$, in
agreement with experimental findings 
\cite{breakup}. 
When the impact velocity $v_0$ is fixed and  the size of the disc $R$
is varied, 
however, the distribution of the $x$ components $n(v_x)$ remains 
the same (see Fig.~\ref{veldistr}(a)). For $n(v_y)$ 
a similar trend is observed as in the case of changing $v_0$, except
that the distribution is less dispersed with varying  $R$
(Fig.~\ref{veldistr}(b)).   
\begin{figure}
\begin{tabular}{cc}
\includegraphics[width=0.49\textwidth]{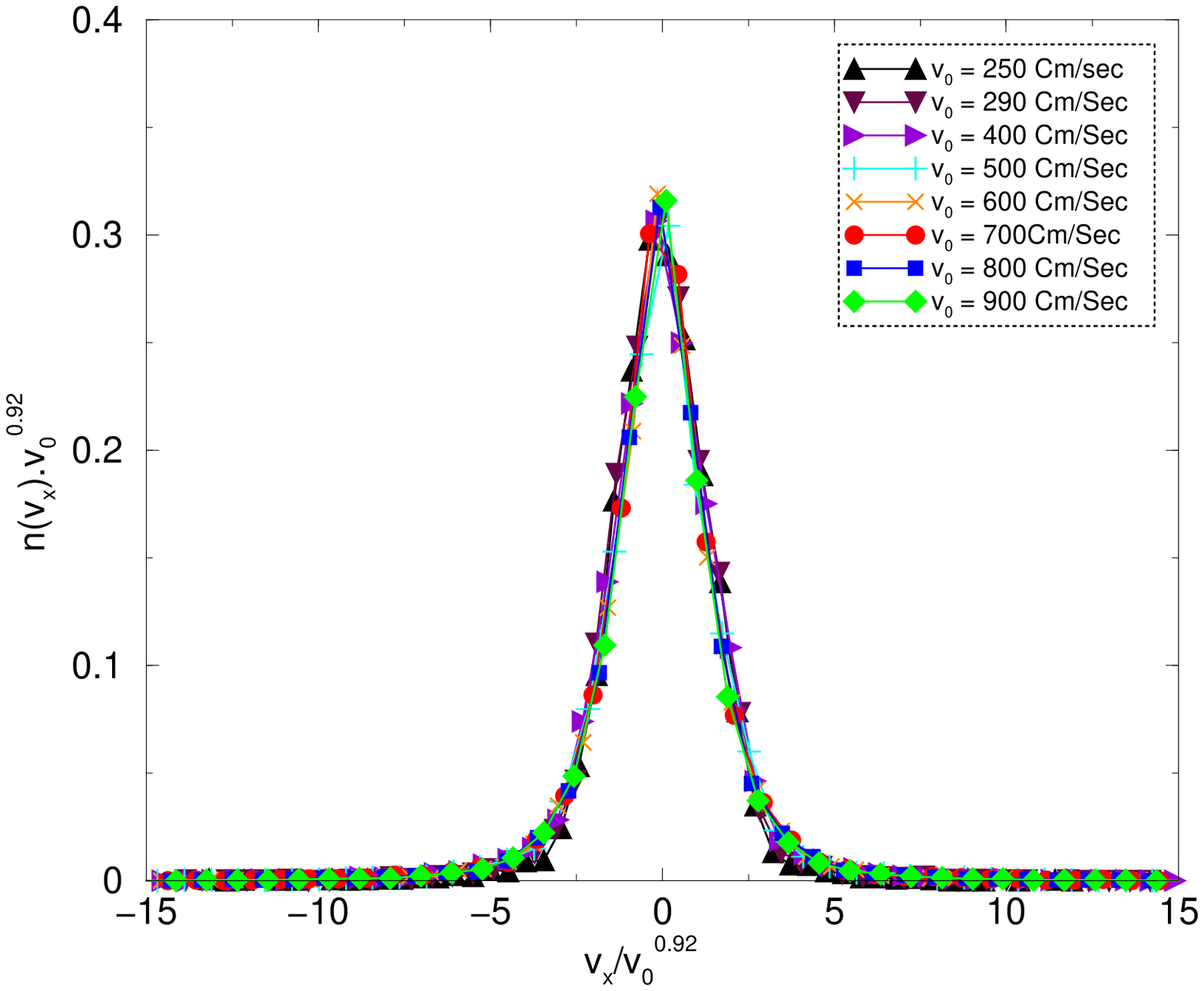}&
\includegraphics[width=0.49\textwidth]{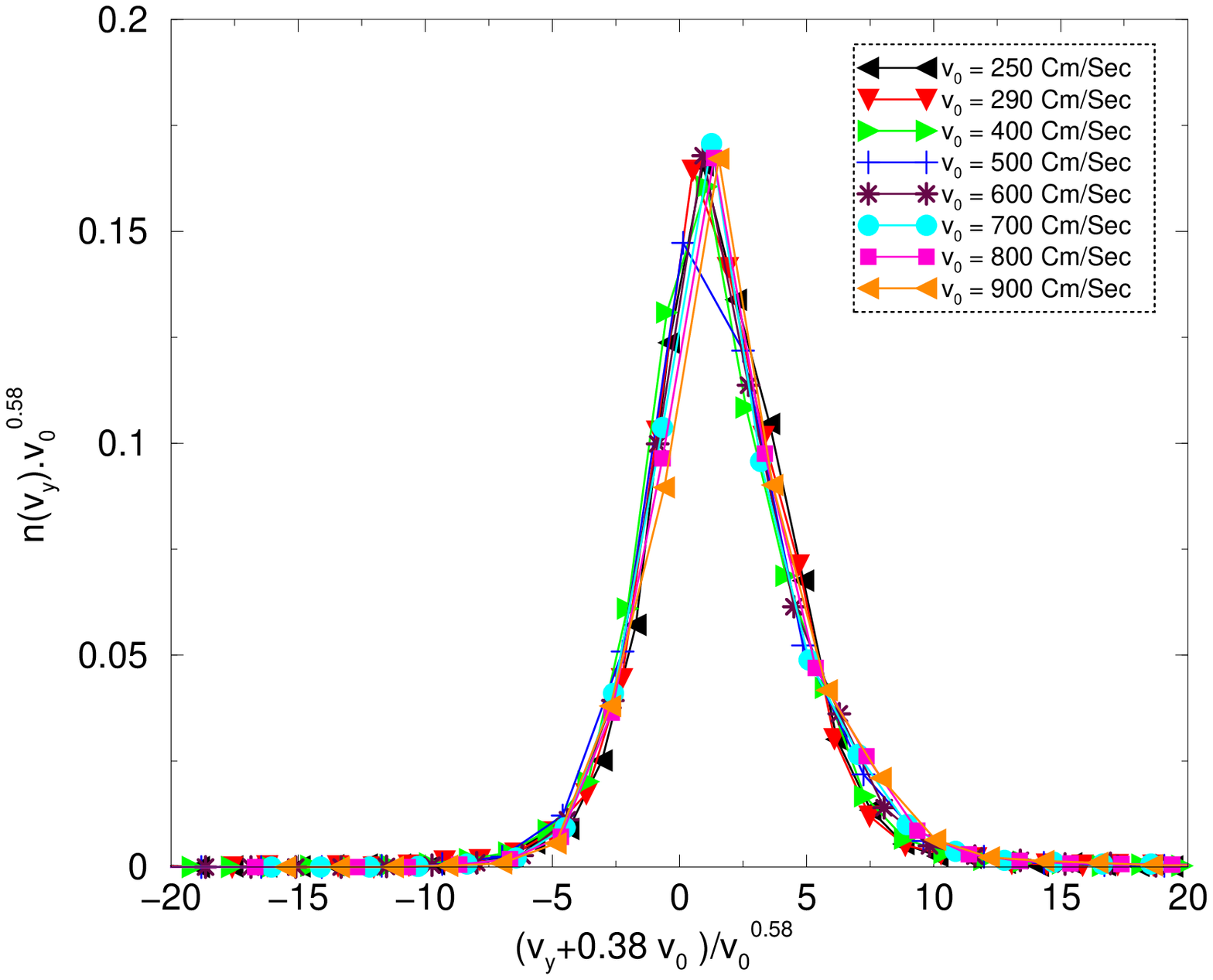}\\
(a)&(b)\\
\end{tabular}
\caption{\label{scalev} Rescaled plots of the velocity distributions 
for a fixed system size $30$ cm varying the impact velocity $v_0$ above
the critical point.}
\end{figure}
\begin{figure}
\begin{center}
\includegraphics[width=0.49\textwidth]{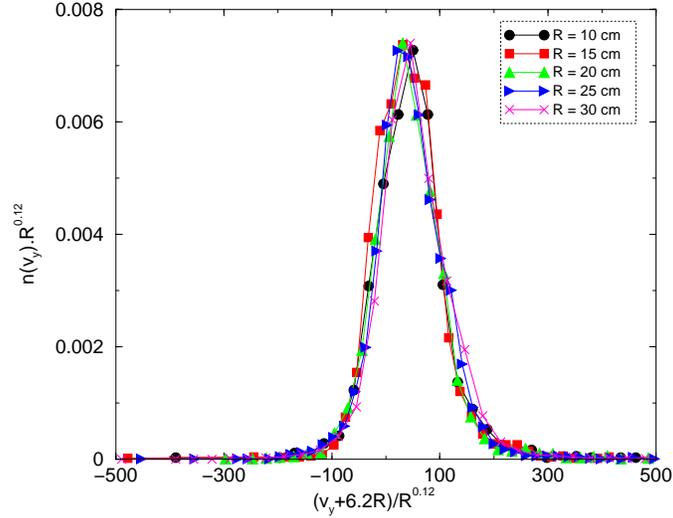}
\end{center}
\caption{ \label{scaler} Rescaled plot of the distributions of the $y$
  component of fragment velocities $n(v_y)$ for a fixed impact
  velocity $400$ cm/sec varying the size $R$ of the particle.
 }
\end{figure}
To reveal the functional form of the dependence of $n(v_x)$ and
$n(v_y)$ on the macroscopic variables $v_0$ and $R$ we performed a
scaling analysis of the distributions.
Figs. \ref{scalev}, \ref{scaler} demonstrate that by appropriate
rescaling the axes one can merge the curves obtained at different
values of the macroscopic variables onto a single curve. In the case
of the $x$ components, the transformation is a stretching and
shrinking by a power of  $v_0$ on the vertical and
horizontal axis, respectively.
However, for the $y$ components,  a combination
of a linear shift and a shrinking by a power of $v_0$ is required.
The good quality of the data collapse implies that the $v_0$ and $R$
dependence of the distributions can be cast in the form
\begin{eqnarray}
\label{eqx}
n(v_x,v_0) &\sim& v_0^{-\alpha} \phi\left(v_x
  v_0^{-\alpha}\right),  \\
\label{eqy}
n(v_y,v_0) &\sim& v_0^{-\beta} \psi\left((v_y+\lambda_1 v_0)
  v_0^{-\beta}\right),   
\end{eqnarray}
where the parameter values $\alpha = 0.92$, $\beta = 0.58$, and
$\lambda_1 = 0.38$ were obtained by varying them until the best data
collapse is obtained in Fig.\ \ref{scalev},\ref{scaler}. Similarly at
constant velocity while varying the 
system size $R$ the functional form of the distribution $n(v_y)$
reads as
\begin{equation}
\label{eqyr}
n(v_y,R) \sim R^{-\gamma} \xi\left((v_y+\lambda_2 R)R^{-\gamma}\right),  
\end{equation}
where $\gamma = 0.12$ and $\lambda_2 = 6.2$ provide the best
collapse in Fig.\ \ref{scaler}. $\phi$, $\psi$, and $\xi$ are scaling
functions which seem to have a Gaussian-like shape in Figs.\
\ref{scalev},\ref{scaler}. The scaling forms 
Eqs.\ (\ref{eqx},\ref{eqy},\ref{eqyr}) show that the width of the
scaling functions have a power law dependence on the impact velocity
$v_0$ and on the radius of the disc $R$. It has to be emphasized that
the above scaling behavior is valid only above the critical velocity
$v_c$; below $v_c$ no scaling was found.

The scaling form of the distribution of the velocity components have
also consequences for the spatial distribution of the flying pieces
after impact. The increase of the width of the distributions with
increasing impact velocity and disc radius implies that the flying
fragments are more dispersed in space. 

\section{Conclusion}
We studied the normal and oblique impact of a circular brittle
particle on a hard frictionless plate using a cell model of cohesive
granular materials. We carried out a detailed analyses of the
evolution of the crack pattern arising in the disc during the impact
process, and of the mass and velocity distributions of fragments in the
final state. 

For both normal and oblique impact, a cone
shaped damage region is formed at the impact point whose base area increases
gradually as the velocity of impact increases. Cracks start to develop
from the the conical damaged region where the maximum stress
concentration exists. 
The oblique crack patterns obtained resemble those of the experimental 
findings \cite{fra_pattern}, where oblique cracks
moving along the plane of maximum compression were found. In agreement
with the experimental 
observations, the oblique cracks in our simulation follow the
trajectory of the maximum compression plane. 
Varying the impact velocities while keeping its normal component constant,
we observed that the crack pattern remains the same in agreement with
recent experimental and theoretical findings
\cite{fra_pattern,breakup,powder,fra_glass,ball_impact}.  

Our analyses showed the existence of a critical value of the impact
velocity, at which the oblique cracks reach  the outer
surface of the disc opposite to the impact point.  The critical
velocity separates two regimes of the break-up process, {\it i.e.} 
below the critical point only a damage cone is formed at the impact
site {\it (damage)}, cleaving of the particle occurs at the critical
point, while above the critical velocity the disc breaks into several
pieces {\it (fragmentation)}. In the limit of very high impact
velocities, the disc suffers complete disintegration into many small
fragments. However, this shattered phase is slowly approached since
the damage ratio and the number of fragments increase logarithmically
with the impact speed. The critical behavior proved to be independent
of the 
impact angle, it solely depends on the normal component of the impact
velocity. Studying the average fragment size revealed that the
critical value of the specific energy increases with the size
of the disc. This implies in practical cases that a higher energy
density is required to break a particle of larger size. 

Above the critical point, the mass distribution $F(m)$ of fragments was
found to obey the Gates-Gaudin-Schuhmann distribution (power law) with
an exponent close to $2$. The power law functional form occurs at
the critical point
and remains unchanged in a broad interval of the impact velocity
independent of the system size and of the impact angle. However, in the
shattered phase attained in the limit of very high impact velocities,
the fragment mass distribution tends to an exponential form. The
results are in good quantitative agreement with 
the experimental findings on the fragmentation of plate-like objects
\cite{glassplate,platelike,composite,ice}.

In applications like mineral processing, a fragment, formed with a
certain velocity, can undergo secondary break-up due to collisions with
other fragments or with the walls of the container. To get an estimate
about the importance of this secondary fragmentation, the study of
fragment velocities is essential. We determined the distribution of
the velocity components of fragments and analysed the scaling
behaviour of the distributions when changing the macroscopic variables
of the system, {\it i.e.} impact velocity $v_0$ amd system size $R$.
A very interesting anomalous scaling of the distribution functions
were revealed with a power law dependence on $v_0$ and $R$.

In a variety of size reduction operations practiced by a wide range 
of industries, there is a particular interest in the net energy required
to achieve a certain size reduction and the energy distribution of 
the fragments during the grinding process. To maximize efficiency of such 
processes, it is important to  know the breakage characteristics of the 
grinding materials. Our current work can provide some of this valuable
information.

\end{document}